\documentstyle[aps,eqsecnum]{revtex}
\begin{document} 
\title{ Renormalization-group approach to the metal-insulator 
transitions in (DCNQI)$_2$M 
(DCNQI is $N,N'$-dicyanoquinonediimine and M=Ag, Cu) }
\author{ K. Yonemitsu } 
\address{ Institute for Molecular Science, Okazaki 444, Japan }
\date{\today}
\maketitle

\begin{abstract}
Metal-insulator transitions and different ground-state phases 
in quasi-one-dimensional materials, ($R_1R_2$-DCNQI)$_2$M 
($R_1$=$R_2$=CH$_3$, I and M=Ag, Cu), are studied with 
a renormalization-group method.
We use one-dimensional continuum models with backward 
scatterings, umklapp processes and couplings with 
2$k_F$ and 4$k_F$ phonons (not static lattice distortion).
We take a quarter-filled band for M=Ag and a sixth-filled band 
coupled with a third-filled band for M=Cu.
Depending on electron-electron and electron-phonon coupling 
strengths, the ground-state phase becomes a Tomonaga-Luttinger 
liquid or a state with a gap(s).
For M=Ag, there appear 
a spin-gap state with a dominant 2$k_F$ charge-density-wave 
correlation, a Mott insulator with a dominant 4$k_F$ 
charge-density-wave correlation, or a spin-Peierls state with 
different magnitudes of spin and charge gaps.
Three-dimensionality is taken into account by cutting off 
the logarithmic singularity in either the particle-particle 
channel or the particle-hole channel.
The difference between the ground-state phase of 
the $R_1$=$R_2$=CH$_3$ salt (spin-Peierls state) and that of 
the $R_1$=$R_2$=I salt (antiferromagnetic state) is 
qualitatively explained by a difference in the cutoff energy 
in the particle-particle channel.
For M=Cu, there appear 
a Mott insulator with a charge density wave of period 3 and 
a Peierls insulator with a charge density wave of period 6.
The conditions for the experimentally observed, Mott insulator 
phase are strong correlation in the sixth-filled band, moderate 
electron-phonon couplings, and finite electron-4$k_F$ phonon 
coupling.
Resistance is calculated as a function of temperature with 
a memory-function approximation in both cases above.
It qualitatively reproduces the differences among the M=Ag and 
M=Cu cases as well as the $R_1$=$R_2$=CH$_3$ and $R_1$=$R_2$=I 
cases.
\end{abstract} 
\pacs{71.30.+h, 71.10.Pm, 72.80.Le, 71.45.Lr}

\section{INTRODUCTION}

Quasi-one-dimensional materials, ($R_1R_2$-DCNQI)$_2$M (M=Ag, Cu), 
show various phases. 
For M=Cu, the hybridization of DCNQI $\pi$ orbitals and Cu $d$ 
orbitals causes intriguing transport\cite{transport} and 
magnetic\cite{magnetic} properties.
The metal-insulator transition is accompanied by the formation of 
a charge density wave (CDW) of period 3.\cite{magnetic,period_3}
The low-temperature phase is a Mott insulator, which is caused by 
strong correlation of $d$ electrons in cooperation with 
electron-lattice coupling and cannot be described by a band picture.
In fact, the spin susceptibility in the insulating state is enhanced 
over the temperature-independent Pauli-like paramagnetism 
seen in the metallic state.
It has called forth theoretical studies, most of which have used 
mean fields\cite{Suzumura} --- the Hartree-Fock approximation 
and the mean field approximation for slave bosons.
The $1/N$ correction to the latter explains 
the overall phase diagram rather well.
The cooperative effect has been discussed on phenomenological 
grounds.\cite{Fukuyama}
Here we take a different approach, which qualitatively reproduces 
exactly known, ground-state properties near the metal-insulator 
transition in one dimension.

For M=Ag, the Ag $d$ level is far away from the Fermi energy so that 
the DCNQI $\pi$ bands are quarter filled.
In this sense, they are similar to (TMTTF)$_2$X and 
(TMTSF)$_2$X salts, which are also quarter-filled and 
show various phases.\cite{IshiguroYamaji}
In the latter, the 4$k_F$ anion potential has been considered 
to produce the umklapp process\cite{EBB} and then to cause 
the metal-insulator transition.\cite{VJE}
Such an extrinsic potential is absent in the ($R_1R_2$-DCNQI)$_2$Ag 
salts, but the metal-insulator transition occurs, accompanied by 
the formation of a 4$k_F$ CDW for both of the 
$R_1$=$R_2$= CH$_3$ (denoted by DMe hereafter) and 
$R_1$=$R_2$=I (denoted by DI hereafter) cases.\cite{Nogami}
At low temperatures, (DMe-DCNQI)$_2$Ag becomes a spin-Peierls state, 
while (DI-DCNQI)$_2$Ag becomes an antiferromagnet.\cite{Hiraki}
Thus electron correlation in the $\pi$ band also plays 
an essential role to determine the ground-state phases.

Electron correlation is essential in low dimensions.
In pure one dimension, the Fermi liquid is unstable against 
any perturbation.
When the perturbation is irrelevant in terms of 
a renormalization group, the excitation spectrum is gapless.
Then the low-energy limit of the property is described by the 
Tomonaga-Luttinger liquid theory and characterized by power laws 
in the density of states and various correlation 
functions.\cite{Solyom,Emery}
Otherwise, the spectrum has a gap so that some of the correlation 
functions decay exponentially.\cite{Luther}

For the metallic (DMe-DCNQI)$_2$Cu, the single-particle spectrum 
has been observed in photoemission experiments and described 
by a power law of the electron binding energy.\cite{Sekiyama}
This suggests that the metallic phase is described by the 
Tomonaga-Luttinger liquid theory for low but not extremely low 
temperatures.
The exponent is much larger than the calculated one for the 
Hubbard model or the extended Hubbard model with on-site 
and nearest-neighbor repulsions only.
However, it can be explained by long-rage 
interactions.\cite{Mila,Capponi}
The power-law holds up to the binding energy as large as 0.3 eV.
This fact would also indicate a long-range interaction.\cite{Sekiyama}
Recently, suppression of the interchain coherent hopping by 
such a large exponent in the single-particle spectrum is numerically 
investigated.\cite{Capponi}

In the insulator phase on the other hand, the Cu $d$ electrons are 
almost localized so that good one-dimensionality is expected.
Since the nesting is perfect and the tendency to an insulator 
phase is strong in one dimension when coupled with phonons, 
the metal-insulator transition must be explained at least 
in a purely one-dimensional model in order for it to occur in 
a quasi-one-dimensional real system.
We then employ a purely one-dimensional model to study 
conditions for the observed Mott insulator phase with a gapless 
spin mode.

Meanwhile, (DCNQI)$_2$Ag salts are insulators for low temperatures. 
One-dimensionality is expected to be better than 
(DCNQI)$_2$Cu salts because the metal ion does not help 
the electron propagation perpendicular to the most conducting 
direction as in the Cu salt.
In fact, the one-dimensional band structure has been evidenced 
by the polarized reflectance spectra on single crystals.\cite{Yakushi}
Thus the Tomonaga-Luttinger liquid theory would be a good 
starting point.
Qualitative aspects of the experimental findings are in fact explained 
by considering the effects of electron-2$k_F$ phonon coupling, 
electron-4$k_F$ phonon coupling, and slight three-dimensionality 
on a one-dimensional continuum model.

In one dimension, a renormalization-group method based 
on the scaling law is very useful to study the effects 
of various perturbations.
The Mott transition is caused by the umklapp process.
The temperature and frequency dependence of the conductivity 
for commensurate and nearly commensurate fillings has been 
studied in detail\cite{Giamarchi} with the renormalization-group 
method combined with a memory-function approximation.\cite{Gotze}
The Mott transition which occurs when the correlation strength 
is changed and that which occurs when the filling is changed have 
different physical properties, which are clarified for even and odd 
commensurabilities.\cite{Schulz_Los_Alamos}
Meanwhile, the electron-phonon interaction produces 
the retarded attraction.
Its effect has also been studied with the renormalization-group method.
When it contributes to the forward scattering, the backward scattering, 
and the umklapp process, it enhances the pairing correlation, 
the formation of a spin gap,\cite{Voit} and 
the formation of a charge gap,\cite{YI,YI2} respectively.

Here, we apply the method to more complex systems, which have 
couplings with 2$k_F$ and 4$k_F$ phonons with one or two bands 
at even or odd commensurability.
We derive and numerically solve the lowest-order equations 
to learn which phase the Tomonaga-Luttinger liquid approaches 
as the energy scale is lowered.
Although a precise description of the electronic states at 
strong-coupling fixed points is beyond the scope of the present method, 
it indicates whether each excitation spectrum has a gap or not 
so that the ground-state phases can be classified accordingly.
Such procedure reproduces qualitative tendency to 
the strong-coupling phases for nearly-half-filled 
electron-phonon systems\cite{YI,YI2} and for 
a two-coupled-chain system.\cite{Fabrizio}
The present paper also follows the procedure.

To compare with experimental results, it would be necessary 
to take weak three-dimensionality into account.
The scaling law and the consequent power-law behavior of 
various quantities in one dimension result from the interference 
of the logarithmic singularity in the 2$k_F$ particle-hole channel 
with that in the particle-particle channel.
The contributions from the corresponding lowest-order bubble 
diagrams are different in signs only.
Quasi-one-dimensionality would be simulated by different 
cutoff energies for the two logarithmic singularities and the 
consequent imbalance between the two channels.\cite{EBB}
In the M=Ag case, we find that the differences between 
the $R_1$=$R_2$=CH$_3$ salt of the spin-Peierls ground state and 
the $R_1$=$R_2$=I salt of the antiferromagnetic ground state 
are qualitatively explained by a difference in the cutoff energy 
in the particle-particle channel.

In the M=Cu case, the two-band feature is essential.
The interference between electron-phonon couplings, 
backward scatterings and umklapp processes is much more 
complicated in the two-band case than in the single-band case.
The phonons with momenta near $2\pi/3$ are responsible 
for the 4$k_F$ scattering in the sixth-filled band and 
for the 2$k_F$ and 4$k_F$ scatterings in the third-filled band, 
while the phonons with momenta near $\pi/3$ are responsible 
for the 2$k_F$ scattering in the sixth-filled band.
Different combinations of the above scatterings lead to different 
backward scattering or umklapp processes, opening charge gaps 
in both of the sixth-filled and third-filled bands.
The insulator phase can have a spin gap only in the third-filled 
band (Mott insulator) or spin gaps in both bands (Peierls insulator).
The former phase is realized when the sixth-filled band ({\it i.e.\/}, 
$\pi$-$d$ hybrid band\cite{Miyazaki}) has strong correlation 
and the electron-2$k_F$ phonon coupling is moderate.
The electron-4$k_F$ phonon coupling is necessary here, 
though it may not be strong.
The latter phase is realized when the electron-2$k_F$ phonon 
coupling is strong.

To study how the behavior of resistance depends on 
the commensurability and model parameters, 
we use a memory-function approximation.
For M=Ag, the renormalization-group method qualitatively reproduces 
the difference between the DMe and DI cases.
For M=Cu, the behavior of the resistance above the transition 
temperature is less sensitive to changes in correlation strengths 
than in the quarter-filled case.
Experimentally, the transition is of first order due to 
the third-order commensurability energy,\cite{third_order} and 
it needs three-dimensionality in phonons, 
which is beyond the scope of the present study.

This paper is organized as follows: 
Sec.~\ref{sec:model} introduces bosonized models 
based on one-dimensional continuum models with 
backward scatterings, umklapp processes, and 
couplings with 2$k_F$ and 4$k_F$ phonons.
Sec.~\ref{sec:renormalization} outlines the derivation of 
renormalization group equations, clarifying how different 
electron-phonon couplings effectively produce different 
backward scattering or umklapp processes and how they 
open gaps in different channels.
Secs.~\ref{sec:results_Ag} and \ref{sec:results_Cu} show 
phase diagrams for different parameters and resistance 
as a function of temperature for M=Ag and M=Cu, respectively, 
to compare them with the experimental results.  
Sec.~\ref{sec:summary} summarizes the present work.
Part of the results presented in this paper were reported 
briefly elsewhere.\cite{Yonemitsu}

\section{MODELS}\label{sec:model}

We consider a continuum model in which the noninteracting part is 
a Tomonaga-Luttinger liquid extended to include the
spin (denoted by subscript $\sigma$) and 
charge (denoted by subscript $\rho$) degrees of freedom,
\begin{equation}
H_{0} =\sum_{\nu=\sigma,\rho} \int \frac{dx}{2\pi} 
\left[ u_\nu K_\nu \left( \partial_x \theta_\nu(x) \right)^2  +
\frac{u_\nu}{K_\nu} \left( \partial_x \phi_\nu(x) \right)^2 \right]
\;,
\end{equation}
where fields $\phi_\nu(x)$ and $(1/\pi)\partial_x \theta_\nu(x)$ 
are conjugate, $u_\nu$ and $K_\nu$ are the velocity and 
the correlation exponent of the $\nu$ channel, respectively, 
which are standard notations and described in detail 
in the previous paper.\cite{YI}
For M=Ag, we consider a quarter-filled band.
For M=Cu, we take two bands and distinguish them by subscripts: 
$A$ for the sixth-filled band; $B$ for the third-filled band.
Then, the noninteracting part is the sum of $H_{A0}$ and $H_{B0}$, 
in which the summation is performed over 
$\nu$=$A\sigma$, $A\rho$ and $\nu$=$B\sigma$, $B\rho$, 
respectively.
It is noted that $H_{(C)0}$ ($C=A,B$) [{\it i.e.\/}, $H_{0}$, 
$H_{A0}$, or $H_{B0}$] contains the (one-electron) kinetic part and 
the forward scattering.
For M=Cu, we consider, for simplicity, the regime where the Fermi 
velocity in the A band and that in the B band do not differ so much.
Since we do not expect that the difference in the Fermi velocities 
affects the scenario to the metal-insulator transition, we use 
the averaged Fermi velocity, $v_F$.
For later convenience, we define 
$X_{(C)\sigma} = 2(1-K_{(C)\sigma}^{-1}) 
\simeq g_{(C)1}/(\pi v_F)$ and  
$X_{(C)\rho} = 2(1-K_{(C)\rho}^{-1}) 
\simeq (g_{(C)1}-2g_{(C)2})/(\pi v_F)$  
for $C=A,B$, where $g_{(C)1}$ and $g_{(C)2}$ are the backward and 
forward scattering strengths (in the C band), respectively.
In terms of the single-band extended Hubbard model 
with on-site ($U$) and nearest-neighbor ($V$) repulsions, 
$X_{ \sigma}= U   /(\pi v_F)$, 
$X_{ \rho}=-(U+4V)/(\pi v_F)$ for quarter filling, 
$X_{A\sigma}=(U+V)/(\pi v_F)$, 
$X_{A\rho}=-(U+3V)/(\pi v_F)$ for sixth filling, and 
$X_{B\sigma}=(U-V)/(\pi v_F)$, 
$X_{B\rho}=-(U+5V)/(\pi v_F)$ for third filling.

\subsection{Electron-electron interactions}

A backward scattering between antiparallel spins 
of strength $Y_{(C)\sigma}= g_{(C)1}/(\pi v_F)$, 
\begin{equation}
H_{(C)\sigma} = Y_{(C)\sigma}\pi v_F\sum_s \int dx \ 
\psi_{(C)1,s}^\dagger(x) \psi_{(C)2,-s}^\dagger(x) 
\psi_{(C)1,-s}(x) \psi_{(C)2,s}(x)
\;,
\end{equation}
where $\psi_{(C)1,s}$ and $\psi_{(C)2,s}$ are 
right- and left-going electrons with spin $s$ (in the $C$ band), 
is written with the phase field as 
\begin{equation}
H_{(C)\sigma} = \frac{Y_{(C)\sigma} v_F}{2\pi\alpha^2} 
\int dx \ \cos \left[2\sqrt{2} \phi_{(C)\sigma}(x)\right]
\;,
\end{equation}
where $\alpha$ is a cutoff parameter of the order of 
the inverse of the Fermi wave number.

To study metal-insulator transition, we need to include 
high-order electron-electron scattering processes, which are 
to be produced in the renormalization process when 
electron-phonon interactions are included.
In the $1/m$-filled band, the Fermi wave number is $\pi/m$ so that 
the umklapp process is written as
\begin{equation}
H_{U,m}[\phi_\rho(,\phi_\sigma);Y_\rho] \propto 
Y_\rho v_F a^{m-2} \int dx 
\left[\sum_s \psi_{1,s}^\dagger(x) \psi_{2,s}(x) \right]^m
+ {\mbox h.c.} 
\;,
\end{equation}
where $Y_\rho$ is a coupling strength such that the prefactor 
appears simple in the phase-field representation, 
$a$ is of the order of the lattice spacing and set to be $\alpha$, 
and the power should be understood as point-split hereafter.
With the phase-field operators, it is rewritten as 
\begin{equation}
H_{U,m}[\phi_\rho(,\phi_\sigma);Y_\rho] = 
\left\{\begin{array}{ll}
( Y_\rho v_F )/(2\pi\alpha^2) 
\int dx \ \cos \left[m\sqrt{2} \phi_\rho(x)\right] 
& \mbox{for even } m \;, \\
( Y_\rho v_F )/(\sqrt{2}\pi\alpha^2) 
\int dx \ \cos \left[m\sqrt{2} \phi_\rho(x)\right]
          \cos \left[ \sqrt{2} \phi_\sigma(x)\right] 
& \mbox{for odd } m \;,
\end{array}\right.
\end{equation}
where and hereafter less relevant terms are neglected.
Thus, we have 
$H_{\rho} = H_{U,4}[\phi_{\rho};Y_{\rho}]$ 
for the quarter-filled band, 
$H_{A\rho} = H_{U,6}[\phi_{A\rho};Y_{A\rho}]$ 
for the sixth-filled band, 
$H_{B\rho\sigma} =
 H_{U,3}[\phi_{B\rho},\phi_{B\sigma};Y_{B\rho\sigma}]$ and 
$H_{B\rho} = H_{U,6}[\phi_{B\rho};Y_{B\rho}]$ 
(which is included though less relevant than $H_{B\rho\sigma}$) 
for the third-filled band.
In terms of the extended Hubbard model, 
$Y_{\rho} = 3U^3/(16\pi^3t^2v_F) = 3Y_\sigma^3/8$, {\it etc.\/}
As $m$ increases, the effect of $ H_{U,m} $ becomes weak 
on opening a gap, as explicitly shown in the 
renormalization-group equations later.
Note that, for odd $m$, the umklapp process involves 
the spin degrees of freedom:\cite{Schulz_Los_Alamos} 
when the umklapp process flows to a strong-coupling fixed point, 
there are gaps in both the charge and spin excitations.
In such a case, soliton excitations carry both charge and spin 
in contrast to the case with even $m$.

In addition, there are interband scattering processes for M=Cu.
The most relevant backward scattering is written as  
\begin{equation}
H_{AB\rho\sigma} = (Y_{AB\rho\sigma} \pi^2 v_F a/\sqrt{2}) 
\int dx \left[\sum_s
\psi_{A1,s}^\dagger(x) \psi_{A2,s}(x) \right]^2
\left[\sum_s
\psi_{B2,s}^\dagger(x) \psi_{B1,s}(x) \right]
+ {\mbox h.c.} 
\;,
\end{equation}
where $Y_{AB\rho\sigma}$ is the coupling strength.
It is rewritten as 
\begin{equation}
H_{AB\rho\sigma} = 
\frac{ Y_{AB\rho\sigma} v_F }{\sqrt{2}\pi\alpha^2} \int dx \ 
\cos \left[2\sqrt{2} \phi_{A\rho}(x)- \sqrt{2} \phi_{B\rho}(x)\right]
\cos \left[\sqrt{2} \phi_{B\sigma}(x)\right]
\;,
\end{equation}
which involves the charge degrees of freedom in the A band and 
both the charge and the spin degrees of freedom in the B band.
The scaling dimension is the lowest (in the noninteracting limit) 
among the perturbations involving the charge degrees of freedom.
So, this is the most likely candidate which causes a metal-insulator 
transition.
When this backward scattering (not umklapp process) flows to a 
strong-coupling fixed point, there are at least three gaps 
in the corresponding channels.
Meanwhile, the most relevant umklapp process is written as  
\begin{equation}
H_{AB\rho} = Y_{AB\rho\sigma} \pi^3 v_F a^2 
\int dx \left[\sum_s
\psi_{A1,s}^\dagger(x) \psi_{A2,s}(x) \right]^2
\left[\sum_s
\psi_{B1,s}^\dagger(x) \psi_{B2,s}(x) \right]^2
+ {\mbox h.c.} 
\;,
\end{equation}
where $Y_{AB\rho}$ is the coupling strength.
It is rewritten as 
\begin{equation}
H_{AB\rho} = \frac{ Y_{AB\rho} v_F }{2\pi\alpha^2}  
\int dx \ 
\cos \left[2\sqrt{2} \phi_{A\rho}(x)+2\sqrt{2} \phi_{B\rho}(x)\right]
\;,
\end{equation}
which involves the charge degrees of freedom in both bands 
but does not involve the spin degrees of freedom.

\subsection{Electron-phonon interactions}

As to phonons, we need to consider those with momenta 
near 2$k_F$ and those with momenta near 4$k_F$.
Because we are not interested in superconductivity here, 
we do not consider phonons with momenta near 0$k_F$, 
though the extension is straightforward.\cite{YI,Yonemitsu}
The wave numbers $\mu$=2$k_F$, 4$k_F$ are 
$\mu$=$\pi/2$, $\pi$ for $m$=4, 
$\mu$=$\pi/3$, $2\pi/3$ for $m$=6, and 
$\mu$=$2\pi/3$, $4\pi/3$(=$-2\pi/3$) for $m$=3.
The corresponding phonon fields and their conjugate momenta are 
denoted by $\phi_{\mu}(x)$ and $\Pi_{\mu}(x)$, respectively.
The fields $\phi_{\pi}(x)$ and $\Pi_{\pi}(x)$ are real, 
and the others are complex.
The phonon parts of the models are written as 
\begin{equation}
H_{\mu} = \int dx \left[ \Pi_{\mu}(x) \Pi_{-\mu}(x)
+\omega_{\mu}^2 \phi_{\mu}(x) \phi_{-\mu}(x) \right]
\;
\end{equation} for the complex fields ($\mu\neq\pi$) and 
\begin{equation}
H_{\pi} = \frac12 \int dx \left[ \Pi_\pi^2(x) 
+ \omega_\pi^2 \phi_\pi^2(x) \right] 
\;
\end{equation} for the real fields, 
where $\omega_{\mu}$ and $\omega_{\pi}$ are the corresponding 
phonon frequencies.

Electron-2$k_F$ phonon coupling is generally written as 
\begin{equation}
H_{1,(C,)\mu}=\sqrt{\pi v_F} \omega_{\mu} y_{(C)1} 
\sum_s \int dx \psi_{(C)2s}^\dagger(x) \psi_{(C)1s}(x) \phi_{\mu}(x) 
+ \mbox{\rm h.c.} 
\;, \end{equation}
where $y_{(C)1}$ is the strength of the 2$k_F$ scattering (in the C band) 
by a phonon.
It is rewritten as 
\begin{equation}
H_{1,(C,)\mu} = 
\frac{\sqrt{v_F} \omega_{\mu} y_{(C)1}}{\sqrt{\pi} \alpha} 
\int dx \ \ \cos \left[ \sqrt{2} \phi_{(C)\sigma}(x) \right] 
e^{-i \sqrt{2} \phi_{(C)\rho}(x)} \phi_{\mu}(x)  + {\mbox h.c.}
\end{equation}
For the quarter-filled band, for example, 
the coupling strength and the phonon frequencies are given by 
$ y_1 = \beta/\sqrt{\pi v_F K} $ and 
$\omega_{\pi/2} = \omega_{\pi} = \sqrt{K/M}$
for the Holstein coupling,\cite{Holstein} 
$ \sum_i \left( \beta q_i n_i + \frac{K}2 q_i^2 + \frac1{2M} p_i^2 \right)
$, with coupling strength $\beta$, spring constant $K$, ionic mass $M$, 
electron density $n_i$ at site $i$, lattice displacement $q_i$ and 
its conjugate momentum $p_i$, and by 
$ y_1 = 2i \alpha_S/\sqrt{\pi v_F K} $ and 
$\sqrt{2}\omega_{\pi/2} = \omega_{\pi} = 2\sqrt{K/M}$
for the SSH coupling,\cite{SSH} 
$ \sum_i \left[ \alpha_S q_{i,i+1} \sum_s 
( c_{i,s}^\dagger c_{i+1,s} + {\mbox h.c.} ) +
\frac{K}{2} q_{i,i+1}^2 + \frac1{2M} p_i^2
\right]
$, with coupling strength $\alpha_S$, $q_{i,i+1} = q_{i+1} - q_{i}$, 
$c_{i,s}$ annihilating an electron with spin $s$ at site $i$, and other 
parameters as defined above.

Electron-4$k_F$ phonon coupling is generally written as 
\begin{equation}
H_{3,(C,)\mu}=-\sqrt{2\pi^3 v_F} \omega_{\mu} y_{(C)3} \sum_{s,s'} a 
\int dx \psi_{(C)2s}^\dagger(x) \psi_{(C)2s'}^\dagger(x) 
\psi_{(C)1s'}(x) \psi_{(C)1s}(x) \phi_{\mu}(x)  
+ \mbox{\rm h.c.}
\;, \end{equation}
where $y_{(C)3}$ is the strength of the 4$k_F$ scattering (in the C band) 
by a phonon.
It is rewritten as 
\begin{equation}
H_{3,(C,)\mu} = 
-\frac{\sqrt{v_F} \omega_{\mu} y_{(C)3}}{\sqrt{2\pi} \alpha}
\int dx \ \ e^{- i 2\sqrt{2} \phi_{(C)\rho}(x)} 
\phi_{\mu}(x)+ {\mbox h.c.}
\label{eq:above_anti}
\end{equation}
For the quarter-filled band, for example, 
the coupling strength is given by 
$ y_3 = \beta'/\sqrt{8\pi^3v_F K} $ 
for the electron-electron-phonon coupling, 
$ \sum_i ( U-\beta' q_i ) n_{i\uparrow} n_{i\downarrow} 
$, with on-site repulsion $U$, coefficient $\beta'$ of its modulation by 
lattice displacement $q_i$, electron density $n_{is}$ with spin $s$ 
at site $i$, and by 
$ y_3 = i \alpha'/\sqrt{2\pi^3v_F K }$
for the electron-electron-phonon coupling, 
$ \sum_{i, s, s'} ( V - \alpha' q_{i,i+1} ) n_{i s} n_{i+1 s'}
$, with nearest-neighbor repulsion $V$, coefficient $\alpha'$ of its 
modulation by lattice displacement $q_{i,i+1}$.

The phonon fields are bilinear, so that they can be integrated out 
completely to produce effective retarded interactions,\cite{YI} 
which are lengthy and not shown here.
In the $\omega_{\mu} \rightarrow \infty $ limit, 
the electron-phonon interactions do nothing but shift the parameters, 
$X_{(C)\sigma} \rightarrow X_{(C)\sigma} - \mid y_{(C)1} \mid^2 $, 
$Y_{(C)\sigma} \rightarrow Y_{(C)\sigma} - \mid y_{(C)1} \mid^2 $, 
$X_{(C)\rho} \rightarrow X_{(C)\rho}
 - \mid y_{(C)1} \mid^2 -4 \mid y_{(C)3} \mid^2$, 
$Y_\rho \rightarrow Y_\rho -(y_3^2 + y_3^{\ast2})/2$, 
$Y_{B\rho\sigma} \rightarrow Y_{B\rho\sigma}
 + (y_{B1} y_{B3} + y_{B1}^\ast y_{B3}^\ast ) $,
$Y_{AB\rho\sigma} \rightarrow Y_{AB\rho\sigma}
 + (y_{B1} y_{A3}^\ast + y_{B1}^\ast y_{A3} )$, and 
$Y_{AB\rho} \rightarrow Y_{AB\rho}
 - (y_{A3} y_{B3} + y_{A3}^\ast y_{B3}^\ast )$.
Below we take the Hamiltonian 
\begin{equation}
H=H_0+H_{\sigma}+H_{\rho}
+H_{\pi/2} +H_{\pi} +H_{1,\pi/2}+H_{3,\pi}
\end{equation}
for M=Ag with a quarter-filled band, and 
\begin{eqnarray}
H=&& \sum_{C=A,B}\left(H_{C0}+H_{C\sigma}+H_{C\rho}\right)
+H_{B\rho\sigma}+H_{AB\rho\sigma}+H_{AB\rho} \nonumber \\
&&+H_{\pi/3}+H_{2\pi/3}
+H_{1,A,\pi/3}+H_{1,B,2\pi/3}+H_{3,A,2\pi/3}+H_{3,B,-2\pi/3}
\end{eqnarray}
for M=Cu with coupled sixth- and third-filled bands.

\section{RENORMALIZATION EQUATIONS}\label{sec:renormalization}

We have derived the equations following the previous study.\cite{YI}
We will give an outline for discussions later.
The present one-dimensional quantum-mechanical system is 
mapped to a two-dimensional classical system where Burgers 
vectors interact with one another.
The correlation function 
\begin{equation}
\langle T_\tau e^{i\sqrt{2} \phi_{(C)\nu} (x,\tau)} 
e^{-i\sqrt{2} \phi_{(C)\nu} (0,0)} \rangle
\ \ \ (C=A,B, \ \ \nu = \sigma, \rho)
\end{equation}
is perturbationally developed and successively integrated by 
changing the length scale little by little.
In this process, there are two possibilities 
for the fate of Burgers vectors: a pair of neutral Burgers vectors 
are annihilated in the larger length scale; 
and a pair of non-neutral Burgers vectors are combined to produce 
another Burgers vector.
The latter is important since it causes various interference 
effects.\cite{YI,YI2}
By comparing the correlation functions in successive length scales, 
we find relations between the effective parameters 
in the larger energy ({\it i.e.\/}, smaller length) scale and those 
in the smaller energy ({\it i.e.\/}, larger length) scale, which are 
described by differential equations (called renormalization-group 
equations) shown below.

\subsection{Quarter-filled band}\label{subsec:Ag}

The combinations for the non-neutral Burgers vectors 
in the renormalization process are 
($y_1$, $y_1^\ast$, $Y_\sigma$), 
($y_3$, $y_3$, $Y_\rho$), and 
($y_3^\ast$, $y_3^\ast$, $Y_\rho$), 
where ($a$, $b$, $c$) means that any two of them are combined 
to produce the complex conjugate of the third.
In order to make a comparison with the half-filled case easier, 
we use the notation similar to that in the previous study,\cite{YI}
$Y_1 = \mid y_1 \mid^2 $, 
$Y_3 = \mid y_3 \mid^2 $, 
and a form factor 
$f = (y_3^2 + y_3^{\ast2})/(2 Y_3)$ 
which satisfies $ -1 < f < 1$.
Below we mark the contributions from the particle-particle channel 
by $J_\eta(l)$ and those from the 2$k_F$ particle-hole channel 
by $J_\kappa(l)$, which are useful to study a three-dimensionality 
effect later.
Then, we finally have 
\begin{equation}
dX_\sigma(l)/dl 
= -J_\eta(l) \left[ X_\sigma(l) - X_\rho(l) \right] X_\sigma(l) /2 
-J_\kappa(l) \left[ X_\sigma(l) + X_\rho(l) \right] X_\sigma(l) /2 
- Y_1(l) D_{\pi/2}(l) 
\;, \label{eq:x_sigma_q}\end{equation}
\begin{equation}
dX_\rho(l)/dl 
= J_\eta(l) \left[ 3 X_\sigma^2(l) + X_\rho^2(l) \right]/4 
-J_\kappa(l) \left\{ \left[3 X_\sigma^2(l) + X_\rho^2(l) \right]/4 
+ 4 Y_\rho^2(l) \right\}
- Y_1(l) D_{\pi/2}(l) -4 Y_3(l) D_\pi(l) 
\;, \label{eq:x_rho_q}\end{equation}

\begin{equation}
dY_\rho(l)/dl = J_\kappa(l) \left[ 2 - 8 K_\rho(l) \right] Y_\rho(l)
- f Y_3(l) D_\pi(l)
\;, \label{eq:y_rho_q}\end{equation}
\begin{equation}
dY_1(l)/dl = J_\kappa(l) \left[ 2 - K_\sigma(l) - K_\rho(l) 
- X_\sigma(l) \right] Y_1(l)
\;, \label{eq:y_1_q}\end{equation}
\begin{equation}
dY_3(l)/dl = J_\kappa(l) \left[ 2 - 4 K_\rho(l) 
- f Y_\rho(l) \right] Y_3(l)
\;, \label{eq:y_3_q}\end{equation}
where $l$=$\ln[E_F/E]$, $E(l)$ is a cutoff energy [$E(0)=E_F$], 
$K_\nu(l) = (1-X_\nu(l)/2)^{-1}$, 
$D_\mu(l)$ is the phonon propagator defined by 
$D_\mu(l) = [\omega_\mu/E(l)] \exp [-\omega_\mu/E(l)]$. 
We have used the relation $X_\sigma(l) = Y_\sigma(l)$ 
due to the spin-rotational symmetry and omitted the equations 
for $ u_\nu(l) $ which does not couple with the above equations 
in the present, lowest order.
Initial conditions at $l=0$ are determined by treating 
the effective retarded interactions carefully.\cite{YI}
For $X_\sigma(0)$, $X_\rho(0)$, and $Y_\rho(0)$, 
only the phonon propagator is integrated from $-\infty$ to
$0$ with the fixed prefactor: $ X_\sigma(0) = X_\sigma 
- Y_1 \left( 1-e^{-\omega_{\pi/2}/E_F} \right)$, for example.

The reason why $X_{\sigma}$ does not renormalize $X_{\sigma}(0)$ or 
why $Y_1$ is not renormalized at $l<0$ in this example has been 
already discussed.\cite{YI}
This correctly reproduces the antiadiabatic limit shown 
below (\ref{eq:above_anti}).

In real materials, slight three-dimensionality would manifest itself 
at very low temperatures where the electron transport 
in the perpendicular direction becomes coherent.
The renormalization-group method would not be justified 
deep in the anisotropic three-dimensional regime, 
where the scaling law no longer holds.
However, the scaling law would deteriorate little by little 
as that regime is approached from above.
Then we can see at least how the renormalization flow is deflected 
and which phase it tends to approach within the present method.
The scaling law in one dimension results from the interference of 
the 2$k_F$ particle-hole channel with the particle-particle channel.
The corresponding lowest-order bubble diagrams are logarithmically 
divergent and have coefficients of equal magnitudes and different 
signs.
Such interference disappears in higher dimensions because the Fermi 
surface does not consist of a finite number of points any more. 
Though the distortion of the Fermi surface removes 
the logarithmic singularity in the particle-hole diagram 
unless the nesting is perfect, 
the particle-particle channel generally becomes less important 
for repulsive interactions when only the particle-particle diagrams 
are summed infinitely.
It is natural to regard quasi-one-dimensionality as causing 
imbalance between the two channels as the most important effect 
among all.\cite{EBB}
It is realized by different cutoff energies for the two logarithmic 
diagrams.

It should be noted that the renormalization-group equations 
derived from the bosonized model and the mapping to 
a two-dimensional classical system and those derived directly 
from the original fermion model are equivalent.\cite{Solyom}
Renormalization of the velocities and detailed forms of cutoff 
functions if any ({\it e.g.\/}, one due to a misfit parameter) 
are generally different, but they are not essential.
Deriving the renormalization equations again in the second method, 
we can distinguish the contributions from the particle-particle 
channel and the contributions from the 2$k_F$ particle-hole channel, 
as already shown above.
Then we can cut off 
either the particle-particle channel by the function $J_\eta(l)$ 
(determined below) 
or the 2$k_F$ particle-hole channel by the function $J_\kappa(l)$.
Assuming that the logarithm $\ln(E_F/E)$ is replaced by 
$\ln[E_F^2/(E^2+\gamma^2 E_F^2)]/2$ ($\gamma=\eta, \kappa$) 
in the perturbation expansions, we obtain 
\begin{equation}
J_\gamma(l) = (d/dl) \ln 
\left[ E_F^2/(E^2(l)+\gamma^2 E_F^2) \right]/2
= (1+\gamma^2 e^{2l})^{-1}
\;,
\end{equation}
which satisfies $J_0(l)=1$, 
$J_{\gamma\neq0}(l)\ll1$ for $l \gg -\ln \gamma$ 
({\it i.e.\/}, $E \ll \gamma E_F$), and 
$ J_{\gamma\neq0}(l) \rightarrow 0$ as $l \rightarrow \infty$.

The temperature dependence of the resistance shows how the system 
approaches the insulating/metallic ground state. 
It is a useful property to characterize the metal-insulator 
transition, and also a good indicator for the quality 
of theoretical approaches.
Since we perform perturbative calculations above, we perform 
the perturbative expansion for the conductivity 
following the study for commensurate and nearly commensurate 
fillings\cite{Giamarchi} with a memory-function formalism. \cite{Gotze} 
The conductivity is given by 
$\sigma(\omega) = (i2u_\rho K_\rho/\pi) 
[\omega + M(\omega)]^{-1}$, 
where the memory function $M(\omega)$ is defined by 
$M(\omega) = \omega \langle j;j \rangle_\omega/
\left(\langle j;j \rangle_0 - \langle j;j \rangle_\omega \right)$ 
and $\langle j;j \rangle_\omega$ is the retarded correlation 
function of the current operator, 
$j(x) = \sqrt{2} u_\rho K_\rho \partial_x \theta_\rho(x) /\pi$.
The lowest-order term in the perturbative expansion 
for $M(\omega)$ is 
$M(\omega) \simeq 
\left(\langle F;F \rangle^0_\omega-\langle F;F \rangle^0_0 \right)/
(-\omega \langle j;j \rangle_0)$,
where $F$ is defined by $F=[j,H]$ and 
$\langle F;F \rangle^0_\omega $ stands for the retarded correlation 
function of the operator $F$ at frequency $\omega$ 
computed in the absence of perturbations.
Later we consider a temperature range below the phonon frequencies, 
where electron-phonon interactions are integrated out.
The calculation of $\langle F;F \rangle^0_\omega $ is 
straightforward.
Taking the $\omega \rightarrow 0$ limit, we finally have, for 
the resistance $R(T) = 1/\sigma(\omega=0,T)$ 
at finite temperature $T$,
\begin{equation}
R(T) \propto Y_\rho^2(T) T 
B^2\left[ 4 K_\rho(T), 1-8 K_\rho(T) \right]
\cos^2 \left[ 4 \pi K_\rho(T) \right]
\;,\end{equation}
where $ B(x)$ is the beta function and 
$Y_\rho(T)$ denotes $Y_\rho[l = \ln(E_F/T)]$, {\it etc.\/}
If we neglect the temperature dependence of $K_\rho$ 
in the equation for $Y_{\rho}$ (and for $\kappa=0$), 
which would be valid only in the weak-coupling limit, 
$R(T) \propto Y_{\rho}^2 
T^{16K_{\rho}-3}$.\cite{Giamarchi} 

\subsection{Coupled sixth- and third-filled bands}\label{subsec:Cu}

The combinations for the non-neutral Burgers vectors 
in the renormalization process are now 
($y_{A3}$, $y_{B1}^\ast$, $Y_{AB\rho\sigma}$), 
($y_{A3}$, $y_{B3}$, $Y_{AB\rho}$), 
($y_{B1}$, $y_{B3}$, $Y_{B\rho\sigma}$), 
($Y_{AB\rho\sigma}$, $Y_{AB\rho}$, $Y_{B\rho\sigma}$), 
($Y_{AB\rho\sigma}$, $Y_{AB\rho\sigma}$, $Y_{B\sigma}$), 
($Y_{B\rho\sigma}$, $Y_{B\rho\sigma}$, $Y_{B\sigma}$), 
($Y_{B\rho\sigma}$, $Y_{B\rho\sigma}$, $Y_{B\rho }$), 
($y_{A1}$, $y_{A1}^\ast$, $Y_{A\sigma}$), and 
($y_{B1}$, $y_{B1}^\ast$, $Y_{B\sigma}$).
When ($a$, $b$, $c$) is possible, 
($a^\ast$, $b^\ast$, $c^\ast$) is also possible 
but it is not listed above because it is trivial.
Note that $y$'s are complex parameters, 
while $Y$'s are real parameters.
Finally, we have 
\begin{equation}
dX_{A\sigma}(l)/dl = - X_{A\sigma}^2(l)
 - \mid y_{A1}(l) \mid^2 D_{\pi/3}(l) 
\;, \label{eq:x_A_sigma}\end{equation}
\begin{equation}
dX_{A\rho}(l)/dl = - Y_{AB\rho\sigma}^2(l) - Y_{AB\rho}^2(l)
 -9 Y_{A\rho}^2(l) - \mid y_{A1}(l) \mid^2 D_{\pi/3}(l)
 -4 \mid y_{A3}(l) \mid^2 D_{2\pi/3}(l) 
\;, \label{eq:x_A_rho}\end{equation}
\begin{equation}
dX_{B\sigma}(l)/dl = - X_{B\sigma}^2(l)
 - \frac14 Y_{AB\rho\sigma}^2(l)
 - \frac14 Y_{B\rho\sigma}^2(l)
 - \mid y_{B1}(l) \mid^2 D_{2\pi/3}(l) 
\;, \label{eq:x_B_sigma}\end{equation}
\begin{equation}
dX_{B\rho}(l)/dl = - \frac14 Y_{AB\rho\sigma}^2(l)
 - Y_{AB\rho}^2(l)  - \frac94 Y_{B\rho\sigma}^2(l) 
 -9 Y_{B\rho}^2(l)  - \left[ \mid y_{B1}(l) \mid^2
  + 4 \mid y_{B3}(l) \mid^2 \right] D_{2\pi/3}(l) 
\;, \label{eq:x_B_rho}\end{equation}
\begin{equation}
dy_{A1}(l)/dl = \left\{ 1 - \frac12
 \left[ K_{A\sigma}(l) + K_{A\rho}(l) + X_{A\sigma}(l)
 \right] \right\} y_{A1}(l)
\;, \label{eq:y_A_1}\end{equation}
\begin{equation}
dy_{A3}(l)/dl = \left[ 1 - 2 K_{A\rho}(l) \right] y_{A3}(l)
 +\frac12 Y_{AB\rho\sigma}(l) y_{B1}(l)
 - \frac12 Y_{AB\rho}(l) y_{B3}^\ast(l)
\;, \label{eq:y_A_3}\end{equation}
\begin{equation}
dy_{B1}(l)/dl = \left\{ 1 - \frac12 
 \left[ K_{B\sigma}(l) + K_{B\rho}(l) + X_{B\sigma}(l)
 \right] \right\} y_{B1}(l) + \frac12 Y_{AB\rho\sigma}(l) y_{A3}(l)
 + \frac12 Y_{B\rho\sigma}(l) y_{B3}^\ast(l)
\;, \label{eq:y_B_1}\end{equation}
\begin{equation}
dy_{B3}(l)/dl = \left[ 1 - 2 K_{B\rho}(l) \right] y_{B3}(l)
 - \frac12 Y_{AB\rho}(l) y_{A3}^\ast(l)
 + \frac12 Y_{B\rho\sigma}(l) y_{B1}^\ast(l)
\;, \label{eq:y_B_3}\end{equation}
\begin{eqnarray}
dY_{AB\rho\sigma}(l)/dl &=&  \left\{ 2 - \frac12
 \left[ 4 K_{A\rho}(l) + K_{B\sigma}(l) + K_{B\rho}(l)
 + X_{B\sigma}(l) \right] \right\} Y_{AB\rho\sigma}(l)
 \nonumber \\
 &-& \frac12 Y_{AB\rho}(l) Y_{B\rho\sigma}(l)
 + \left[ y_{B1}(l) y_{A3}^\ast(l) + 
        y_{B1}^\ast(l) y_{A3}(l) \right] D_{2\pi/3}(l)
\;, \label{eq:y_AB_rho_sigma }\end{eqnarray}
\begin{eqnarray}
dY_{AB\rho}(l)/dl &=&  \left\{ 2 - 2 \left[
 K_{A\rho}(l) + K_{B\rho}(l) \right] \right\} Y_{AB\rho}(l)
 \nonumber \\
 &-& \frac12 Y_{AB\rho\sigma}(l) Y_{B\rho\sigma}(l)
 - \left[ y_{A3}(l) y_{B3}(l)
       + y_{A3}^\ast(l) y_{B3}^\ast(l) \right] D_{2\pi/3}(l)
\;, \label{eq:y_AB_rho }\end{eqnarray}
\begin{eqnarray}
dY_{B\rho\sigma}(l)/dl &=&
 \left\{ 2 - \frac12 \left[
 K_{B\sigma}(l) + 9 K_{B\rho}(l) + X_{B\sigma}(l)
 + Y_{B\rho}(l) \right] \right\} Y_{B\rho\sigma}(l)
 \nonumber \\
 &-& \frac12 Y_{AB\rho\sigma}(l) Y_{AB\rho}(l)
 + \left[ y_{B1}(l) y_{B3}(l) + y_{B1}^\ast(l) y_{B3}^\ast(l)
 \right] D_{2\pi/3}(l)
\;, \label{eq:y_B_rho_sigma}\end{eqnarray}
\begin{equation}
dY_{A\rho}(l)/dl =  \left[ 2 - 18 K_{A\rho}(l) \right] Y_{A\rho}(l)
\;, \label{eq:y_A_rho}\end{equation}
\begin{equation}
dY_{B\rho}(l)/dl =  \left[ 2 - 18 K_{B\rho}(l) \right] Y_{B\rho}(l)
 -  \frac14 Y_{B\rho\sigma}^2(l)
\;, \label{eq:y_B_rho}\end{equation}
where the cutoff energy and the phonon propagators are defined 
as before, 
$X_{C\sigma}(l) = Y_{C\sigma}(l)$ ($C=A,B$) has been used.
Initial conditions at $l=0$ are also determined as before.

For the present system, the resistance $R(T)$ in a temperature range 
below the phonon frequencies is given by 
\begin{eqnarray}
R(T) &\propto& 
    5 Y_{AB\rho\sigma}^2(T) T \cos^2
     \left[\left\{4 K_{A\rho}(T)+ K_{B\rho}(T)
       + K_{B\sigma}(T)\right\} 
 \pi/4 \right] \nonumber \\ & \times & 
B^2\left[ \left\{4 K_{A\rho}(T)+ K_{B\rho}(T)
       + K_{B\sigma}(T)\right\}/4, 
        1-\left\{4 K_{A\rho}(T)+ K_{B\rho}(T)
       + K_{B\sigma}(T)\right\}/2
     \right] \nonumber \\
&+&8 Y_{AB\rho}^2(T) T \cos^2
     \left[\left\{K_{A\rho}(T)+ K_{B\rho}(T)\right\} 
 \pi \right] 
B^2\left[ \ K_{A\rho}(T)+ K_{B\rho}(T), 
        1-2\left\{K_{A\rho}(T)+ K_{B\rho}(T)\right\}
     \right] \nonumber \\
&+&9 Y_{B\rho\sigma}^2(T) T \cos^2
     \left[\left\{9 K_{B\rho}(T)+ K_{B\sigma}(T)\right\} 
 \pi/4 \right] 
B^2\left[ \left\{9 K_{B\rho}(T)+ K_{B\sigma}(T)\right\}/4, 
        1-\left\{9 K_{B\rho}(T)+ K_{B\sigma}(T)\right\}/2
     \right] \nonumber \\
&+&36 Y_{A\rho }^2(T) T \cos^2
     \left[9K_{A\rho}(T) \pi \right] 
B^2\left[9 K_{A\rho}(T), 1-18 K_{A\rho}(T) \right] \nonumber \\
&+&36 Y_{B\rho }^2(T) T \cos^2
     \left[9 K_{B\rho}(T) \pi \right] 
B^2\left[9 K_{B\rho}(T), 1-18 K_{B\rho}(T)
     \right] 
\;.\end{eqnarray}
In the above formula, any perturbation which contains $\phi_{A\rho}$ 
or $\phi_{B\rho}$ contributes to the resistance.
Among them, $Y_{AB\rho\sigma}$ has the lowest scaling dimension 
and dominates the resistance.
In the weak-coupling limit, we have 
$R(T) \propto Y_{AB\rho\sigma}^2 
T^{4K_{A\rho}+K_{B\sigma}+K_{B\rho}-3}$. 
The actual temperature dependence is modified from this 
simple power law.

\section{RESULTS FOR Quarter-filled band}\label{sec:results_Ag}

The fixed points of the spin and charge correlation exponents, 
$K_\sigma^\ast$ and $K_\rho^\ast$, determine 
the asymptotic correlation functions.  
Assuming a gap in the $\nu$ spectrum for $K_\nu^\ast $ = 0, 
we classify the ground-state phases into 
a gapless state in which all perturbations are irrelevant so that 
the low-energy limit is regarded as a Tomonaga-Luttinger liquid 
(denoted by TL in the phase diagrams below), 
a state with only a spin gap which we call a spin-gap state 
(denoted by SG), 
a state with only a charge gap which we call a Mott insulator 
(denoted by MI), and 
a state with both spin and charge gaps which we call a spin-Peierls 
state (denoted by SP).
When we consider three-dimensionality, the gapless (thus metallic) 
state does not correspond to a Tomonaga-Luttinger liquid any more 
so that it is simply denoted by M. 
The Tomonaga-Luttinger liquid has a dominant 2$k_F$ SDW correlation 
$\sim \exp[-(K_\rho^\ast+1)\ln r]$ or a dominant 4$k_F$ CDW 
correlation $\sim \exp[-4 K_\rho^\ast  \ln r]$. 
The spin-gap state has a dominant 2$k_F$ CDW correlation 
$\sim \exp[-  K_\rho^\ast  \ln r]$. 
The Mott insulator has a dominant 4$k_F$ CDW correlation.

Usually a spin-Peierls state is described by a localized 
spin system since the charge gap is much larger than the spin gap.
Here we call even a state with comparable magnitudes of gaps 
a spin-Peierls state because it is smoothly connected 
with the so-called spin-Peierls state.
We assume that phonons are also one-dimensional for simplicity 
so that we do not consider the possibility for a static lattice 
distortion.
In any case, a coupling with 2$k_F$ phonons is a necessary condition 
here for a spin gap and thus for a spin-Peierls state 
because we consider repulsive electron-electron interactions only.
With a coupling with three-dimensional phonons, the spin-Peierls 
state here would be accompanied by a static lattice distortion.

Before we use parameters for (DCNQI)$_2$Ag, we first show how 
the electron-4$k_F$ phonon coupling and the umklapp process 
interfere with each other and next show how they are affected by 
the electron-2$k_F$ phonon coupling, large phonon frequency and 
nearest-neighbor repulsion.
For the bare phonon dispersion, we take 
$\omega_{\pi/2}$ = $\omega_{\pi} / \sqrt{2}$.
It is noted that the quantity $f$ appears with $Y_\rho(l)$ 
as a prefactor in the renormalization equations so that 
we fix $f=-1$ and allow $Y_\rho$ to be negative.
We do not cut off the logarithmic singularities, 
$\kappa$ = $\eta$ = $0$, unless explicitly mentioned.

\subsection{Coupling with 4$k_F$ phonons and umklapp process}

It is noted again that the perturbative expansion gives 
$Y_\rho = 3X_\sigma^3/8$ for the extended Hubbard model.
The umklapp process is of third order with respect to 
electron-electron interaction, so that it is generally weak.
In this and the next subsections, we vary the strength of 
the umklapp process to see the interference effect.

To have a charge gap, $K_\rho^\ast$ must vanish so that 
$X_\rho(l)$ must approach $-\infty$ in the $l \rightarrow 
\infty$ limit.
From the renormalization equation (\ref{eq:x_rho_q}), it requires 
$Y_\rho(l)$ not to vanish.
Then from (\ref{eq:y_rho_q}), $K_\rho(l)$ must be smaller than $1/4$ 
[$X_\rho(l)$ must be smaller than $-6$] for large $l$.
For the pure Hubbard model, it is known that it cannot be 
smaller than $1/2$.\cite{Schulz}
Even for the extended Hubbard model with nearest-neighbor repulsion, 
it is at least $3/16$.\cite{Mila}
Such a small $K_\rho(l)$ is achieved generally for long-range 
repulsion.\cite{Schulz_long_range}
The photoemission experiment\cite{Sekiyama} may indeed suggest a 
long-range repulsion for (DMe-DCNQI)$_2$Cu.
In the (DCNQI)$_2$Ag salts, the 4$k_F$ CDW is 
observed\cite{Nogami} and the lattice would be modulated.
So, it is reasonable to regard electron-phonon interaction 
as cooperating with the umklapp process to have such a small 
value of $K_\rho(l)$.
This is the case when the umklapp process itself is 
very weak because it occurs as a high-order process.

In the absence of the electron-phonon coupling, the critical value 
for a finite charge gap is large (Fig.~\ref{fig:phaseAg01}) 
and beyond the scope of the perturbative regime.
However, with the electron-4$k_F$ phonon coupling, it becomes 
smaller.
For $Y_\rho>0$ corresponding to $f=-1$, {\it e.g.\/}, 
for lattice modulation of the nearest-neighbor repulsion, 
the umklapp process is constructively interfered with 
the electron-4$k_F$ phonon coupling, as is expected from the 
renormalization-group equations.
This is the reason for the negative slope of 
the critical value of $Y_3$ at $Y_\rho=0$ in the figure.
It is noted that, even for $Y_\rho<0$ corresponding to $f=+1$, 
{\it e.g.\/}, for lattice modulation of the on-site repulsion, 
the critical value of $Y_3$ does not become so large.
This is because $Y_\rho(l)$ changes the sign after it decreases 
due to the destructive interference at the initial high-energy scale.
Thus, the interference is constructive at low-energy scales 
irrespectively of the sign of $f$.
This situation is in contrast to the half-filled and 
nearly-half-filled cases, where the opening of a spin gap is 
sensitive to the form factor of the electron-2$k_F$ phonon 
coupling.\cite{YI,YI2}

\subsection{Coupling with 2$k_F$ phonons}

The electron-2$k_F$ phonon coupling is expected to decrease 
$X_\rho(l)$, enhancing the tendency for a charge gap, from 
(\ref{eq:x_rho_q}).
However, its tendency is very weak as clearly seen from 
a comparison of Fig.~\ref{fig:phaseAg01} with Fig~\ref{fig:phaseAg02}
and directly from Fig.~\ref{fig:phaseAg03}.
There is no interference of the electron-2$k_F$ phonon coupling 
with the umklapp process in (\ref{eq:y_rho_q}).
This indicates again how the interference effect is important 
to cause the metal-insulator transition.

The role of the electron-2$k_F$ phonon coupling is mainly to 
enhance the tendency for a spin gap.
For a sufficiently large coupling, the system has a finite spin gap 
(Fig.~\ref{fig:phaseAg03}) as expected from (\ref{eq:x_sigma_q}).
Then, the 2$k_F$ CDW correlation becomes dominant.
Note the phase boundary between TL and SG and that between 
MI and SP depends considerably on the electron-4$k_F$ phonon 
coupling.
This is due to the fact that the electron-2$k_F$ phonon coupling 
is affected by both of the spin and charge correlation exponents in 
(\ref{eq:y_1_q}).
Meanwhile, the electron-4$k_F$ phonon coupling is affected by only 
the charge correlation exponent in (\ref{eq:y_3_q}).
Even in the weak-coupling limit [$K_\sigma(l) \simeq K_\rho(l) 
\simeq 1$], $Y_1(l)$ is largely affected by the correlation 
exponents as shown in (\ref{eq:y_1_q}), although $Y_3(l)$ always 
decreases in (\ref{eq:y_3_q}).
Thus, the opening of a spin gap is rather sensitive to 
the electron-4$k_F$ phonon coupling.

\subsection{Phonon frequency and nearest-neighbor repulsion}

Now we have a clear idea about how the electron-phonon couplings 
interfere with the electron-electron interaction and how they 
determine the ground-state phase.
In this subsection, we show how the phase diagram depends on 
the phonon frequency and the relative magnitude of 
$\mid X_\rho \mid$ to $X_\sigma$, {\it i.e.\/}, the 
nearest-neighbor repulsion in terms of the extended Hubbard model.
Note that $- X_\rho / X_\sigma = 1 + 4 V/U$ in this model.
For $V/U=1/4$ and $1/2$, the relative magnitude of $\mid X_\rho \mid$ 
becomes twice and three times as large as that for the Hubbard model, 
respectively, so that the results can be largely affected by 
the nearest-neighbor repulsion strength.

The phonon frequency relative to the Fermi energy controls 
the energy scale of the interference.
In the antiadiabatic limit, the electron-phonon coupling simply 
shifts the strengths of the electron-electron scattering parameters.
For finite phonon frequency below the Fermi energy, 
the effective interaction is retarded and the interference effect becomes 
largest at an energy scale comparable to the phonon frequency 
as in the half-filled and nearly-half-filled systems.\cite{YI,YI2}
As the phonon frequency increases, the critical coupling strength 
of $Y_3$ for opening a charge gap decreases, while that of $Y_1$ 
for opening a spin gap increases [Figs.~\ref{fig:crossAg04}(a) and (b)].
Then the Mott insulator phase occupies a wider area in the phase 
diagram.

As the magnitude of $\mid X_\rho \mid$ increases, the initial value 
of $K_\rho$ decreases.
Then the critical value of $Y_3$ for opening a charge gap decreases 
[Fig.~\ref{fig:crossAg04}(c)].
The slope is steeper for smaller phonon frequencies.
Meanwhile, the $Y_1$ coordinate of the cross section of the phase 
boundaries among TL, MI, SG, and SP does not 
depend so much on $\mid X_\rho \mid$.

\subsection{Phase diagram for (DCNQI)$_2$Ag}

According to the local density functional theory\cite{Miyazaki}, 
the band width for (DMe-DCNQI)$_2$Ag is about 0.9eV.
Phonons due to the dimerization are observed at about 0.08eV
in the infrared spectrum\cite{Meneghetti}.  We use 
$\sqrt{2} \omega_{\pi/2}$ = $\omega_{\pi}$ = $0.4E_F$.
The condition for the Mott insulator phase is easily achieved 
because the phonon frequency relative to the Fermi energy is 
not small and because the nearest-neighbor repulsion relative to 
the on-site repulsion is not small either.\cite{Kanoda}
We show the data\cite{Yonemitsu} again for completeness 
(Fig.~\ref{fig:phaseAg05}).

The author does not know the electron-electron and electron-phonon 
interaction strengths, but the qualitative aspect of the present results 
would not change for different electron-electron interaction strengths.
Since (DMe-DCNQI)$_2$Ag becomes a spin-Peierls state at low 
temperatures (in the sense that both of the spin and charge gaps open),
we expect that its electron-phonon coupling strengths are in the 
SP phase of the figure.
Note that all of the four possible phases appear in the phase diagram 
in the purely one-dimensional case.
This is because the Tomonaga-Luttinger liquid phase is stable for 
weak electron-phonon couplings.
We consider a cutoff in the logarithmic singularity below, where the 
Tomonaga-Luttinger liquid is not realized.
Then some phases disappear from the phase diagram.

\subsection{Effects of a cutoff in the logarithmic singularity}

We consider three-dimensionality by cutting off 
the logarithmic singularity in either the particle-particle 
channel or the 2$k_F$ particle-hole channel as explained 
in the previous section.
The most important effect appears in the charge degrees of 
freedom.
It is obvious from (\ref{eq:x_rho_q}) that a cutoff in the 
particle-particle channel, $\eta \neq0$, leads $K_\rho^\ast$ 
to 0, while a cutoff in the 2$k_F$ particle-hole channel, $\kappa \neq0$, 
leads $K_\rho^\ast$ to infinity. 
Therefore, a charge gap opens in the former case, while it does not open 
in the latter case.
Then, possible phases are a Mott insulator and a spin-Peierls state 
in the former case (Fig~\ref{fig:phaseAg06}), and a gapless metallic state 
and a spin-gap state in the latter case (Fig~\ref{fig:phaseAg07}).

If the nesting property is maintained in the quasi-one-dimensional case,
the particle-particle channel becomes less important at low energies.
Then, it is reasonable that the nesting causes a finite charge gap or 
makes the charge gap larger.
Note that the phase boundary for a finite spin gap is shifted to the right 
in Fig.~\ref{fig:phaseAg06}, as expected from (\ref{eq:x_sigma_q}).
As the cutoff increases, a transition occurs from the spin-Peierls state 
to the Mott insulator at zero temperature.
This is consistent with the experimental results.
(DMe-DCNQI)$_2$Ag has strong anisotropy so that it is regarded as 
a good one-dimensional material and it becomes an insulator below 
about 120 to 150K.
Meanwhile, (DI-DCNQI)$_2$Ag has a considerable conductivity in the 
transverse direction and it is an insulator already at room temperature.
From the activation plot, the charge gap is estimated to be 
490K\cite{Hiraki}.
(DMe-DCNQI)$_2$Ag becomes a spin-Peierls state at about 80K 
so that it has a finite spin gap at zero temperature.
Meanwhile, (DI-DCNQI)$_2$Ag becomes antiferromagnetic below 
5.5K\cite{Hiraki} so that the spin excitation spectrum is gapless.
The Mott insulator in the present study is expected to become 
an antiferromagnetic [or spin-density-wave (SDW)] state when weak 
three-dimensionality is taken into account because the repulsive 
interaction would produce an effective antiferromagnetic coupling 
in the transverse direction.
The above result is reminiscent of the earlier work for localized spins 
at half filling which showed the instability of the spin-Peierls state 
in quasi-one dimension against a SDW.\cite{Inagaki}

If the nesting property is lost in the quasi-one-dimensional case, 
the possibility for a finite charge gap would be very strong electron 
correlation, which is beyond the scope of the perturbative 
renormalization-group approach.
This possibility may not be excluded, but we can say at least whether 
the temperature dependence of the resistance within the present 
approach is qualitatively consistent with the experimental data or not.
It will be done in the next subsection.
The weakened tendency for a spin gap is common with both 
cutoffs.
Note that the tendency for a spin gap is substantially suppressed here 
(Fig.~\ref{fig:phaseAg07}, $\kappa\neq0$) compared with the other case 
(Fig.~\ref{fig:phaseAg06}, $\eta\neq0$).
This is reasonable in that, in the three-dimensional case free from 
nesting ($\kappa\neq0$), a spin gap opens 
when the phonon-mediated effective attraction 
overcomes the repulsion, while such condition is not necessary for 
the spin gap in a spin-Peierls state.

\subsection{Temperature dependence of the resistance}

Our previous work\cite{Yonemitsu} has shown results which 
are not yet converged so that we show the convergent results here.
Here, temperatures are in the unit of $E_F \simeq 2400$K.
When the particle-particle channel is cut off, the resistance 
increases and the metal-insulator transition temperature 
increases (Fig.~\ref{fig:resistanceAg08}).
With a slight three-dimensional component, the resistance increases 
rapidly below the transition temperature.
This is obtained by renormalizing both $Y_\rho$ and $K_\rho$.
If only $Y_\rho$ is renormalized, the resistance is not 
so steep below the transition temperature.
Including this, the overall and qualitative behavior of the resistance 
is indeed consistent with the experimental one for (DMe-DCNQI)$_2$Ag 
and (DI-DCNQI)$_2$Ag.\cite{Hiraki}
Thus, we consider that a cutoff in the particle-particle channel 
imitates the effect of three-dimensionality for (DCNQI)$_2$Ag salts.
When the 2$k_F$ particle-hole channel is cut off on the other hand, 
the resistance decreases and the system finally becomes metallic 
at low temperatures (Fig.~\ref{fig:resistanceAg09}).
This is obviously contradictory with the experimental data, as it 
is expected from the renormalization-group equations.

The real difference between (DMe-DCNQI)$_2$Ag 
and (DI-DCNQI)$_2$Ag would not be limited to the difference 
in the three-dimensionality.
In fact, (DI-DCNQI)$_2$Ag has smaller bandwidth\cite{Miyazaki} 
so that the relative strength of electron-electron interaction would 
be larger than (DMe-DCNQI)$_2$Ag.
We calculated the resistance with increasing coupling strengths 
and found that the behavior is similar to the case with increasing 
cutoffs in the particle-particle channel (not shown).
However, the spin degrees of freedom is almost unaffected.
Therefore, the three-dimensionality is the key to understand their 
differences, although the relative coupling strength is also different 
and contributes to the difference in the resistance.

\section{RESULTS FOR Coupled sixth- and third-filled 
bands}\label{sec:results_Cu}

The ground-state phases can be classified according to 
the channel(s) whose excitation spectrum has a gap.
When any perturbation is irrelevant and vanishes 
in the low-energy limit, the limit is described as a 
Tomonaga-Luttinger liquid without any gap.
When all of the four $X_\nu$'s diverge and both bands 
have spin and charge gaps, the system would be basically 
described as a band insulator (if the spin and charge gaps 
are comparable in the magnitude).
It would be realized if a CDW of period 6 is formed 
and the Brillouin zone is folded at $\pm\pi/6$ so that both 
bands have a gap at the Fermi points.
Then, it may be called a Peierls insulator 
because it can occur by the Peierls mechanism only, {\it i.e.\/},  
without the help of electron correlation.
The experimentally observed, insulator phase have 
a CDW of period 3.
There are charge gaps in both bands and a spin gap in the 
third-filled band only.
In fact, it has a long-ranged antiferromagnetic order 
at low temperature.
Since it can not be described by a band picture, we call it 
a Mott insulator.
In the phase diagrams below, each phase is denoted by 
the number of gap(s).
The three important phases above, the Tomonaga-Luttinger liquid, 
the Mott insulator, and the Peierls insulator, are denoted by 
0, 3, and 4, respectively.

There are also other phases which do not generally occupy 
a wide area in the phase diagrams below.
They can be artifacts of the present, lowest-order 
renormalization-group approach if the occupied area is 
very narrow.
The mechanism of each phase is discussed below.
The phase 1 has a spin gap either in the B band 
(if the correlation in the A band is comparable to or stronger 
than that in the B band) or in the A band (if the correlation 
in the B band is much stronger).
The former would correspond to a freely moving (unpinned) 
CDW of period 3.
The phase 2 has spin gaps in both bands (if $y_1 \neq 0$)
or charge gaps in both bands (if $y_1 = 0$).
The former would correspond to a freely moving, 
CDW of period 6.

\subsection{Interference between different interactions}

Before looking at the numerical results, we should consider what 
is suggested by the equations.
When the fixed point of the correlation exponent 
$K_\nu^\ast$ is zero, there is a gap in the $\nu$ channel.
For that, $X_\nu(l)$ must diverge to $-\infty$ as $l$ goes to $\infty$ 
in the corresponding equation among 
(\ref{eq:x_A_sigma}) -- (\ref{eq:x_B_rho}).
Since the phonon propagators exponentially decrease and thus 
the electron-phonon interactions ($y$'s) are finally integrated out, 
the behavior of the electronic perturbations, $Y_{AB\rho\sigma}(l)$, 
$Y_{AB\rho}(l)$, $Y_{B\rho\sigma}(l)$, $Y_{A\rho}(l)$, and 
$Y_{B\rho}(l)$, is of particular interest.
If some of them diverge, they make the corresponding $X_\nu(l)$ 
diverge to $-\infty$ in Eqs.~(\ref{eq:x_A_sigma}) -- (\ref{eq:x_B_rho}).
The behavior of each perturbation $Y(l)$ largely depends upon 
the factor multiplied by $Y(l)$ on the right-hand side of $dY(l)/dl$ 
[for example, $2 - \frac12  \left[ 4 K_{A\rho}(l) + K_{B\sigma}(l) 
+ K_{B\rho}(l) + X_{B\sigma}(l) \right]$ for $Y_{AB\rho\sigma}(l)$].
Note that, before the equations for $Y_{A\sigma}(l)$ and 
$Y_{B\sigma}(l)$ (not shown) are linearized to retain 
the spin-rotational symmetry, 
the factors were $2 - 2 K_{A\sigma}(l)$ for $Y_{A\sigma}(l)$ 
and $2 - 2 K_{B\sigma}(l)$ for $Y_{B\sigma}(l)$.
If we neglect the interference effect, {\it i.e.\/}, if we do not consider the 
renormalization process where a pair of non-neutral Burgers vectors 
are combined to produce another Burgers vector, these factors are 
given by ($l$ dependence is implicit hereafter) 
$2 - \frac12(4K_{A\rho}+K_{B\sigma}+K_{B\rho})$ 
for $Y_{AB\rho\sigma}$, 
$2 - 2 ( K_{A\rho} + K_{B\rho} )$ for $Y_{AB\rho}$, 
$2 - \frac12 ( K_{B\sigma} + 9 K_{B\rho} )$ for $Y_{B\rho\sigma}$, 
$2 - 18 K_{A\rho}$ for $Y_{A\rho}$, 
$2 - 18 K_{B\rho}$ for $Y_{B\rho}$, 
$2 - 2 K_{A\sigma}$ for $Y_{A\sigma}$, and 
$2 - 2 K_{B\sigma}$ for $Y_{B\sigma}$.
They are 2 $-$ ``scaling dimensions'' in the field-theoretical terminology.
As the scaling dimension becomes lower, the perturbation generally 
becomes more relevant.
In the weak-coupling limit ($X_\nu \rightarrow 0$ and 
$K_\nu \rightarrow 1$), they are 
$-1$ for $Y_{AB\rho\sigma}$, 
$-2$ for $Y_{AB\rho}$, 
$-3$ for $Y_{B\rho\sigma}$, 
$-16$ for $Y_{A\rho}$ and $Y_{B\rho}$, and 
$0$ for $Y_{A\sigma}$ and $Y_{B\sigma}$.
The corresponding factors for the electron-phonon interactions are 
$0$ for $y_{A1}$ and $y_{B1}$ and $-1$ for $y_{A3}$ and $y_{B3}$.
Namely, $Y_{A\sigma}$, $Y_{B\sigma}$, $y_{A1}$, and $y_{B1}$ 
are marginal and the others are irrelevant in this limit.
Then, a charge gap would not open in either band.

Of course, the relevance depends upon the correlation strengths.
As $K_\nu$'s become smaller, the perturbations become more relevant.
More importantly, the interference effects largely affect the relevance 
of each perturbation in the renormalization process.
The strengths of $Y_{AB\rho\sigma}$, $Y_{AB\rho}$, 
$Y_{B\rho\sigma}$, $Y_{A\rho}$, and $Y_{B\rho}$ 
are initially very small (set to be zero in the numerical calculations 
in the next section) because they correspond to high-order 
electron-electron scattering processes.
However, the electron-$2\pi/3$ phonon interactions, 
$y_{A3}$, $y_{B1}$, and $y_{B3}$, interfere with one another to 
produce $Y_{AB\rho\sigma}$, $Y_{AB\rho}$, $Y_{B\rho\sigma}$, 
and the latter three then interfere with one another as well as with 
the electron-$2\pi/3$ phonon interactions.
Once $Y_{AB\rho\sigma}$ becomes relevant and diverges, 
$X_{A\rho}$, $X_{B\sigma}$, and $X_{B\rho}$ diverges to $-\infty$ 
and gaps open in the corresponding channels.
If $Y_{AB\rho}$ becomes relevant, gaps open in the channels 
$A\rho$ and $B\rho$.
If $Y_{B\rho\sigma}$ becomes relevant, gaps open in the channels 
$B\sigma$ and $B\rho$.
Since the scaling dimension of $Y_{AB\rho\sigma}$ is the lowest 
among the three, the first situation would always occur when the second 
or the third situation is realized.
Without the electron-4$k_F$ phonon interactions, $y_{A3}$ and $y_{B3}$, 
on the other hand, $Y_{AB\rho\sigma}$, $Y_{AB\rho}$, 
$Y_{B\rho\sigma}$, and $Y_{B\rho}$ remain zero 
(when they are initially set to be zero), 
so that a charge gap does not open in either band.
In such a case, the situation for $y_{B1}$ and $X_{B\sigma}$ is the same 
as that for $y_{A1}$ and $X_{A\sigma}$ discussed below.
In contrast to $Y_{AB\rho\sigma}$, $Y_{AB\rho}$, $Y_{B\rho\sigma}$, 
and $Y_{B\rho}$, the perturbation $Y_{A\rho}$ does not interfere with 
any other perturbation.
Its scaling dimension is much higher than the others so that it would 
not play an important role.

It is noted that, if $Y_{AB\rho\sigma}$ or $Y_{B\rho\sigma}$ is finite 
(though it may be very small in the real materials), it and $y_{B1}$ would 
interfere and produce $y_{A3}$ or $y_{B3}$, respectively.
This is also a possibility for opening charge gaps if an infinitesimal (or 
very small) value of $y_{A3}$ or $y_{B3}$ is necessary in the numerical 
results with a vanishing (initial value of) $Y_{AB\rho\sigma}$ or 
$Y_{B\rho\sigma}$ in the next section.
In the present two-band system, it is essential that phonons with 
momenta near $2\pi/3$ contribute to $y_{A1}$, $y_{A3}$, and $y_{B3}$ 
processes.
In the single-band quarter-filled case, the coupling with 2$k_F$ phonons 
interferes with the backward scattering, and the coupling with 4$k_F$ 
phonons interferes with the umklapp process, but the two couplings 
do not interfere with each other(Sec.~\ref{subsec:Ag}) in contrast to 
the present case.

The electron-$\pi/3$ phonon interaction $y_{A1}$ does not interfere with 
another electron-phonon interaction.
It only interferes with $X_{A\sigma}$.
After $y_{A1}$ is integrated out, {\it i.e.\/}, 
$\mid y_{A1} \mid^2 D_{\pi/3}$ vanishes at $l>l_0$ 
for some $ l_0$, $X_{A\sigma}$ follows $X_{A\sigma}(l) = 
[l-l_0+X_{A\sigma}^{-1}(l_0)]^{-1}$.
The fixed point is either $X_{A\sigma}^\ast = 0$ 
[$K_{A\sigma}^\ast=1$] 
or $X_{A\sigma}^\ast = -\infty$ [$K_{A\sigma}^\ast=0$], depending 
upon the sign of $X_{A\sigma}(l_0)$.
In other words, when the electron-$\pi/3$ phonon interaction $y_{A1}$ 
is strong enough for $X_{A\sigma}(l_0)$ to be negative, a gap opens in 
the channel $A\sigma$.
Otherwise, the excitation spectrum of this channel is gapless.
If the electron-4$k_F$ phonon interactions, $y_{A3}$ and $y_{B3}$, are 
absent, the behavior of $y_{B1}$ and $X_{B\sigma}$ is exactly the same 
as that of $y_{A1}$ and $X_{A\sigma}$ mentioned above.
Without the electron-2$k_F$ phonon interactions, $y_{A1}$ and $y_{B1}$, 
$Y_{AB\rho\sigma}$ and $Y_{B\rho\sigma}$ remain zero 
(when they are initially set to be zero), 
so that a spin gap does not open in either band for the repulsive case, 
where the initial conditions for $X_{A\sigma}$ and $X_{B\sigma}$ are 
positive.
In short, electron-2$k_F$ phonon coupling is necessary for a spin gap, 
while electron-4$k_F$ phonon coupling is necessary for a charge gap 
unless a sufficiently strong long-range interaction makes the scaling 
dimension of a high-order electron-electron scattering process 
lower than 2.
This statement holds also for the single-band quarter-filled 
case.(Sec.~\ref{sec:results_Ag})

\subsection{Phase diagrams}

In the phase diagrams below, the charge gaps in the phases 
3 and 4 are brought about by $Y_{AB\rho\sigma}$, while 
the spin gap in the A band in the phases 2 and 4 are 
brought about by $y_{A1}$.
The spin gap in the B band is caused by either 
$Y_{AB\rho\sigma}$ or $y_{B1}$.
We initially set $y_{A1} = y_{B1} = y_1$ and 
$y_{A3} = y_{B3} = y_3$ and vary $y_1$ and $y_3$ 
for several sets of 
($X_{A\sigma}$, $X_{A\rho}$,
 $X_{B\sigma}$, $X_{B\rho}$), where 
$X_{A\rho}$ and $ X_{B\rho}$ are negative.
The strengths of $Y_{AB\rho\sigma}$, $Y_{AB\rho}$, 
$Y_{B\rho\sigma}$, $Y_{A\rho}$, and $Y_{B\rho}$ 
are initially ({\it i.e.\/}, at $l=-\infty$) set to be zero 
in the numerical calculations since they are 
of high order with respect to electron-electron scatterings.
For the bare phonon dispersion, we take 
$2 \omega_{\pi/3}$ = $2 \omega_{2\pi/3} /\sqrt{3}$ = $0.4E_F$
as in the previous section unless explicitly 
mentioned.

When electron correlation is weak in both bands 
(Fig.~\ref{fig:phaseCu01}), the phase diagram mainly consists of 
the phase 0 (Tomonaga-Luttinger liquid) and the phase 4 (Peierls 
insulator), as expected, unless only the electron-4$k_F$ phonon 
coupling $y_3$ is strong.
As the electron-2$k_F$ phonon coupling $y_1$ increases, 
a spin gap opens in the B band either mainly by $y_{B1}$ 
(phases $0\rightarrow1$) or by constructive interference 
of $y_{B1}$ with $y_{A3}$ and $y_{B3}$ (phases $0\rightarrow3$).
The phase 1 may be changed into the phase 3.
As $y_1$ further increases, a spin gap opens also in the A band 
(phases $1\rightarrow2$, phases $3\rightarrow4$).
It should be noted that the correlation strength ($X_{A\sigma}$, 
$X_{A\rho}$), the electron-2$k_F$ phonon coupling 
strength ($y_{A1}$), and the electron-4$k_F$ phonon coupling 
strength ($y_{A3}$) in the A band are the same as the corresponding 
coupling strengths in the B band in this figure, but a spin gap 
opens first in the B band.
This is due to the constructive interference of $y_{B1}$ with 
$y_{A3}$ and $y_{B3}$.
In fact, the critical coupling strengths for spin gaps are the 
same for $y_3 = 0$ and different for $y_3 \neq 0$.
As $y_3$ increases, the difference becomes large because 
the critical coupling strength for a spin gap in the B band 
becomes small.

When electron correlation is strong in the A band 
(Fig.~\ref{fig:phaseCu02}), 
a stronger electron-2$k_F$ phonon coupling $y_1$ 
is necessary for a spin gap in the A band, as expected.
If the correlation remains weak in the B band, 
the relation between the critical coupling strength 
for a spin gap in the B band and that for charge gaps 
is very similar to the weakly correlated case.
Namely, as $y_1$ increases, a spin gap opens in the B band 
either mainly by $y_{B1}$ (phases $0\rightarrow1$) or 
by constructive interference of $y_{B1}$ with 
$y_{A3}$ and $y_{B3}$ (phases $0\rightarrow3$).
Then, the phase 1 is changed into the phase 3.
Finally, when $y_1$ overcomes the strong correlation 
in the A band, a spin gap opens in the A band 
(phases $3\rightarrow4$).
As a consequence, the phase diagram mainly consists of 
the phases 0 (Tomonaga-Luttinger liquid), 3 (Mott insulator 
with a CDW of period 3), and 4 (Peierls 
insulator with a CDW of period 6).
It should be noted again that the electron-4$k_F$ phonon 
coupling is necessary for charge gaps.
On the $y_1$ axis ($y_3 = 0$), the possible phases are 0, 1, and 2.
Therefore, the condition for the phase 3 is strong 
correlation in the A band, moderate electron-phonon 
couplings (not too strong to overcome the electron 
correlation in the A band), and finite electron-4$k_F$ phonon 
coupling though it may be small.

On the other hand, when electron correlation is strong 
in the B band (Fig.~\ref{fig:phaseCu03}), 
a stronger electron-2$k_F$ phonon coupling $y_1$ 
is necessary for a spin gap in the B band, unless 
$y_3$ is strong.
Meanwhile, the critical coupling strength for 
a spin gap in the A band remains small if the 
correlation remains weak in the A band.
This situation does not correspond to (DCNQI)$_2$Cu 
salts because the correlation is expected to be 
stronger in the A band.
If $y_3$ is small, a spin gap in the A band opens first 
and then the other three gaps open as $y_1$ increases.
If $y_3$ is large on the other hand, the three gaps 
except the spin gap in the A band open first.
Note the phase 1 here is different from that 
in the other figures in the sense that the A band 
has a spin gap here.

When electron correlation is strong in both bands 
and their strengths are the same 
(Fig.~\ref{fig:phaseCu04}), the phase 1 disappears 
and the phase 2 exists only if $y_3 = 0$ (spin gaps) 
or if $y_1 = 0$ (charge gaps).
The critical coupling strength for a spin gap 
in the B band is the same as that in the A band 
for $y_3 = 0$ and becomes smaller than the latter 
once $y_3 \neq 0$ as in Fig.~\ref{fig:phaseCu01}.
The two curves for these critical coupling strengths 
are shifted to the right for small $ y_3 $, compared with 
Fig.~\ref{fig:phaseCu01}.

So far we took $X_{A\sigma} = \mid X_{A\rho} \mid$ and 
$X_{B\sigma} = \mid X_{B\rho} \mid$, assuming only the 
on-site repulsion in electron-electron interactions.
When we consider nearest-neighbor repulsion, for example, 
such a relation no longer holds.
Recall that, in terms of the single-band extended Hubbard model, 
$X_{A\sigma}=(U+V) /(\pi v_F)$, 
$X_{A\rho}=-(U+3V) /(\pi v_F)$ for sixth filling, and 
$X_{B\sigma}=(U-V) /(\pi v_F)$, 
$X_{B\rho}=-(U+5V) /(\pi v_F)$ for third filling.
Even for the same coupling strengths in terms of 
the extended Hubbard model, the scattering strengths 
depend on the filling factor.
Neglecting the difference in the Fermi velocities 
for simplicity and taking the $X$ values 
of the single-band extended Hubbard model 
for sixth and third fillings, 
we study the effect of the nearest-neighbor 
repulsion with $U /(\pi v_F)=0.4$ and $V /(\pi v_F)=0.1$ 
(Fig.~\ref{fig:phaseCu05}).
The resultant phase diagram is similar to 
Fig.~\ref{fig:phaseCu02} because $X_{A\sigma} > X_{B\sigma}$.
When compared with the result for $U /(\pi v_F)=0.4$ 
and $V /(\pi v_F)=0$ (Fig.~\ref{fig:phaseCu01}), 
the nearest-neighbor repulsion increases the critical coupling strength 
for a spin gap in the A band and decreases that for a spin gap in the B 
band, as expected from the fact that the nearest-neighbor repulsion 
increases $X_{A\sigma}$ and decreases $X_{B\sigma}$.

Experimentally, pressure induces the metal-insulator transition.
The transition is of first order due to the third-order commensurability 
energy\cite{third_order} and it needs three-dimensionality in phonons, 
which is beyond the scope of the present study.
As temperature decreases, the resistance abruptly increases at the 
transition under pressure for ($R_1R_2$-DCNQI)$_2$Cu 
with $R_1$=$R_2$=CH$_3$ or I, but above the transition temperature 
the resistance behaves as at ambient pressure where no transition 
occurs.
The pressure would change the band width at least, but it hardly 
affects the behavior of the resistance above the transition temperature.
We decrease the band width by multiplying $X$'s, $\mid y \mid^2$'s, 
and $\omega$'s (all of which scale as the inverse of the band width) 
by a common factor (1.1 to 1.4) and calculated the temperature 
dependence of the resistance 
in the case of Fig.~\ref{fig:phaseCu02} with $y_1$ and $y_3$ near the 
boundary between the phases 0 and 3 (Fig.~\ref{fig:resistanceCu06}).
The overall behavior of the resistance is insensitive to changes 
in the parameters even near the metal-insulator transition, 
which is qualitatively consistent with the experimental results.
This is in contrast to the quarter-filled case appropriate for 
(DCNQI)$_2$Ag salts, where the correlation strength 
affects the behavior of the resistance rather 
sensitively.(Sec.~\ref{sec:results_Ag})

\section{SUMMARY}\label{sec:summary}

Metal-insulator transitions and electronic phases in (DCNQI)$_2$M 
(M=Ag, Cu) salts are studied with the renormalization-group 
method for the one-dimensional continuum models with 
backward scatterings, umklapp processes and 
couplings with 2$k_F$ and 4$k_F$ phonons.
These salts are in contrast to the quarter-filled (TMTTF)$_2$X and 
(TMTSF)$_2$X salts, where the extrinsic 4$k_F$ anion potential 
produces the umklapp process.
For the present salts with M=Ag, such a potential is absent 
but the electron-4$k_F$ phonon coupling interferes constructively with 
the umklapp process, thereby causing a metal-insulator transition.
The 4$k_F$ CDW is therefore a product of the cooperation 
of the electron-electron and electron-phonon interactions.
It can be viewed as a Mott insulator in the sense that it has gapless 
spin excitations and it cannot be described by a band picture.

Experimentally, the physical properties of ($R_1R_2$-DCNQI)$_2$Ag 
depend upon $R_1R_2$.
For $R_1$=$R_2$=CH$_3$, one-dimensionality is rather good.
It becomes first an insulator at about 120K and then a 
spin-Peierls state at 80K, opening a spin gap.
Meanwhile, for $R_1$=$R_2$=I, electron transfer in the perpendicular 
direction is not negligible and electron correlation is expected 
to be stronger due to the narrower bandwidth.
It is already an insulator at room temperature and the spin 
excitation spectrum remains gapless to zero temperature.
In fact, it becomes antiferromagnetic below 5.5K.
Such qualitative difference is explained by the present approach 
if the three-dimensionality is taken into account by cutting off 
the logarithmic singularity in the particle-particle channel.
The cutoff suppresses the opening of a spin gap and enhances 
the charge gap.
The stronger correlation for $R_1$=$R_2$=I would contribute to 
further enhancing the charge gap. 
The temperature dependence of the resistance is calculated with 
a memory-function approximation and its behavior is consistent 
with the experimentally observed one.

For M=Cu with coupled sixth- and third-filled bands, 
the Mott transition is accompanied by the formation of a CDW.
In real materials, three-dimensionality would not be neglected.
Reasons why we take the one-dimensional model are: 
a Tomonaga-Luttinger behavior is observed in the metallic phase 
in photoemission experiments; and the metal-insulator transition 
should be explained in one dimension also.

In order for gaps to open in the charge excitations in both bands and in 
the spin excitation in the third-filled band, the high-order backward 
scattering $Y_{AB\rho\sigma}$ must be relevant.
Its scaling dimension is higher than 2 in the noninteracting limit.
So, it may look as an irrelevant perturbation.
It is not the case if the interference of the electron-2$k_F$ phonon 
coupling $y_{B1}$ with the electron-4$k_F$ phonon couplings, $y_{A3}$ 
and $y_{B3}$, is taken into account.
These electron-phonon couplings interfere with one another and produce 
effective high-order backward and umklapp scatterings.
The interference of the electron-phonon couplings and the high-order 
scatterings is constructive and can make $Y_{AB\rho\sigma}$ 
a relevant perturbation.

In order for a gap not to open in the spin excitation in the sixth-filled 
band, the electron correlation must be strong enough.
Therefore, the condition for the experimentally observed, Mott insulator 
phase with a CDW of period 3, {\it i.e.\/}, with a gapless 
spin mode, is strong correlation in the sixth-filled, $\pi$-$d$ hybrid band, 
moderate electron-phonon couplings which are not too strong to overcome 
the strong electron correlation in the sixth-filled band, and finite 
electron-4$k_F$ phonon coupling which may be small.
The temperature dependence of the resistance is found to be insensitive 
to changes in the parameters even near the metal-insulator transition, 
which is again consistent with the experimentally observed one.
If electron-phonon couplings were too strong, the insulator phase would 
be accompanied by a CDW of period 6 and have gaps in all channels.

Showing these results, we have demonstrated that 
the renormalization-group method reproduces the qualitative aspects 
of the ground-state phases and the behavior of the resistance in these 
quasi-one-dimensional organic materials very well.

\acknowledgments
The author would like to thank A. Chainani, K. Hiraki, and K. Kanoda
for enlightening discussions and showing him data prior to publication 
and H. Fukuyama for useful comments.
This work was supported by a Grant-in-Aid for Scientific Research 
on Priority Areas 
``Anomalous Metallic State near the Mott Transition'' and 
``Novel Electronic States in Molecular Conductors'' 
from the Ministry of Education, Science, Sports and Culture.

\begin{figure}\caption{
Phase diagram for 
$X_\sigma$ = $0.4$, $X_\rho$ = $-0.8$, 
$\omega_{\pi}$ = $0.1E_F$, and $Y_1 = 0$.
}\label{fig:phaseAg01}\end{figure}

\begin{figure}\caption{
Phase diagram for 
$X_\sigma$ = $0.4$, $X_\rho$ = $-0.8$, 
$\omega_{\pi}$ = $0.1E_F$, and $Y_1 = 0.2$.
}\label{fig:phaseAg02}\end{figure}

\begin{figure}\caption{
Phase diagram for 
$X_\sigma$ = $0.4$, $X_\rho$ = $-0.8$, 
$Y_\rho$ = $0.375 X_\sigma^3$, and $\omega_{\pi}$ = $0.1E_F$.
}\label{fig:phaseAg03}\end{figure}

\begin{figure}\caption{
$Y_1$ and $Y_3$ coordinates of the cross section of the phase 
boundaries among TL, MI, SG, and SP, 
(a) as a function of $\omega_{\pi} / E_F$ for $X_\rho$ = $-0.8$, 
(b) as a function of $\omega_{\pi} / E_F$ for $X_\rho$ = $-1.2$, and 
(c) as a function of $\mid X_\rho \mid$ for $\omega_{\pi}$ = $0.1E_F$.
The other parameters are the same as in Fig.~ 3.
}\label{fig:crossAg04}\end{figure}

\begin{figure}\caption{
Phase diagram for 
$X_\sigma$ = $0.4$, $X_\rho$ = $-0.8$, 
$Y_\rho$ = $0.375 X_\sigma^3$, and 
$\omega_{\pi}$ = $0.4E_F$.
}\label{fig:phaseAg05}\end{figure}

\begin{figure}\caption{
Phase diagram with a cutoff in the particle-particle channel, 
$\eta$ = $0.1$.
The other parameters are the same as in Fig.~5.
}\label{fig:phaseAg06}\end{figure}

\begin{figure}\caption{
Phase diagram with a cutoff in the particle-hole channel, 
$\kappa$ = $0.1$.
The other parameters are the same as in Fig.~5.
}\label{fig:phaseAg07}\end{figure}

\begin{figure}\caption{
Logarithm of resistance (in arbitrary unit) as a function of 
temperature (in the unit of $E_F$) with different cutoffs in the 
particle-particle channel, 
$\eta$ = 0, 0.02, 0.04, 0.06, 0.08, and 0.1 from the bottom. 
The parameters are $Y_1 = 0.25$, $Y_3 = 0.6$, and otherwise 
the same as in Fig.~5.
}\label{fig:resistanceAg08}\end{figure}

\begin{figure}\caption{
Logarithm of resistance (in arbitrary unit) as a function of 
temperature (in the unit of $E_F$) with different cutoffs in the 
particle-hole channel, 
$\kappa$ = 0, 0.01, 0.02, 0.03, 0.04, and 0.05 from the top.
The parameters are $Y_1 = 0.4$, $Y_3 = 0.8$, and otherwise 
the same as in Fig.~5.
}\label{fig:resistanceAg09}\end{figure}

\begin{figure}\caption{
Phase diagram for 
($X_{A\sigma}$, $X_{A\rho}$, $X_{B\sigma}$, $X_{B\rho}$) = 
(0.4, $-$0.4, 0.4, $-$0.4).
}\label{fig:phaseCu01}\end{figure}

\begin{figure}\caption{
Phase diagram for 
($X_{A\sigma}$, $X_{A\rho}$, $X_{B\sigma}$, $X_{B\rho}$) = 
(0.8, $-$0.8, 0.4, $-$0.4).
}\label{fig:phaseCu02}\end{figure}

\begin{figure}\caption{
Phase diagram for 
($X_{A\sigma}$, $X_{A\rho}$, $X_{B\sigma}$, $X_{B\rho}$) = 
(0.4, $-$0.4, 0.8, $-$0.8).
}\label{fig:phaseCu03}\end{figure}

\begin{figure}\caption{
Phase diagram for 
($X_{A\sigma}$, $X_{A\rho}$, $X_{B\sigma}$, $X_{B\rho}$) = 
(0.8, $-$0.8, 0.8, $-$0.8).
}\label{fig:phaseCu04}\end{figure}

\begin{figure}\caption{
Phase diagram for 
($X_{A\sigma}$, $X_{A\rho}$, $X_{B\sigma}$, $X_{B\rho}$) = 
(0.5, $-$0.7, 0.3, $-$0.9).
}\label{fig:phaseCu05}\end{figure}

\begin{figure}\caption{
Logarithm of resistance (in arbitrary unit) as a function of 
temperature (in the unit of $E_F$) for different correlation strengths.
The parameters are ($X_{A\sigma}$, $X_{A\rho}$,
 $X_{B\sigma}$, $X_{B\rho}$) = 
($0.8$, $-0.8$, $0.4$, $-0.4$), $y_1 = 0.4$, $y_3 = 0.3$, 
and $\omega$'s as before for the curve at the bottom.
For the other curves, the parameters 
$X$'s, $\mid y \mid^2$'s, and $\omega$'s 
are 1.1, 1.2, 1.3, and 1.4 times the above values.
}\label{fig:resistanceCu06}\end{figure}

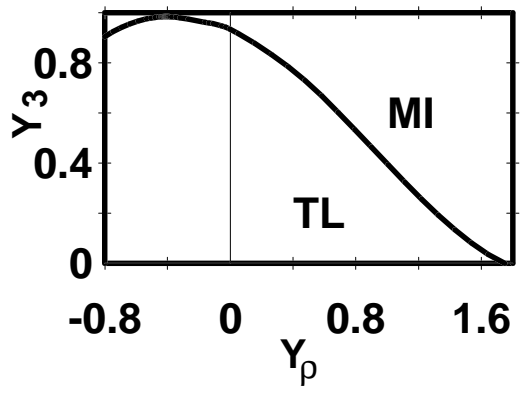
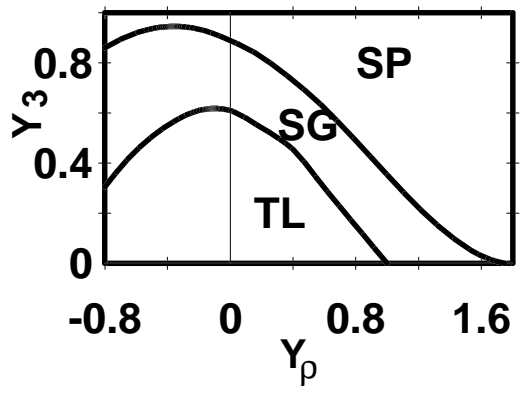
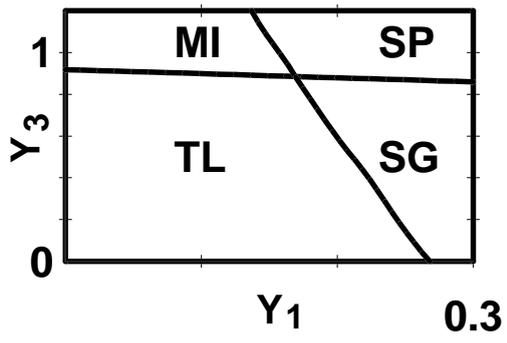
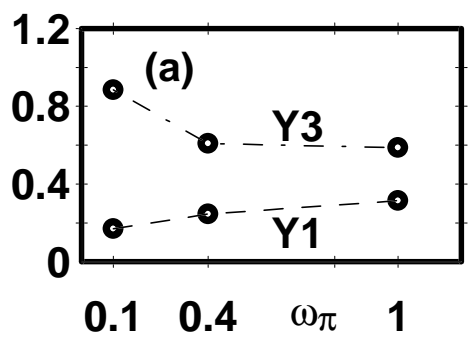
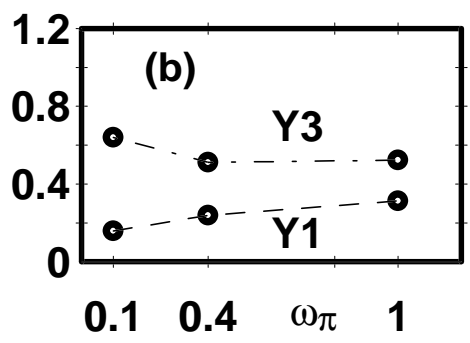
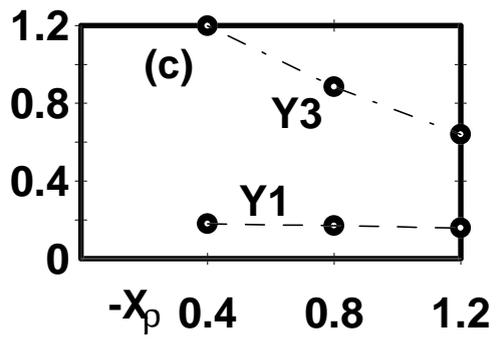
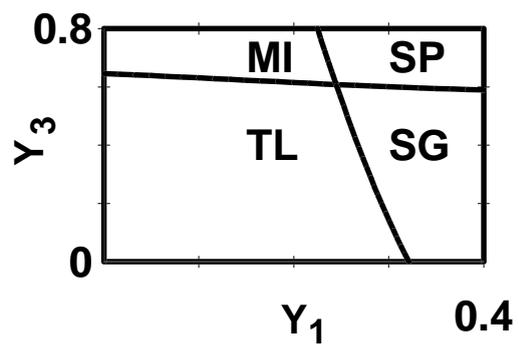
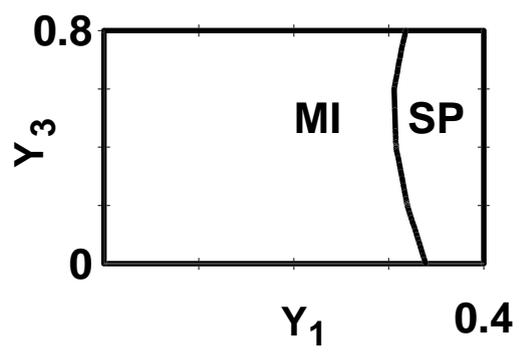
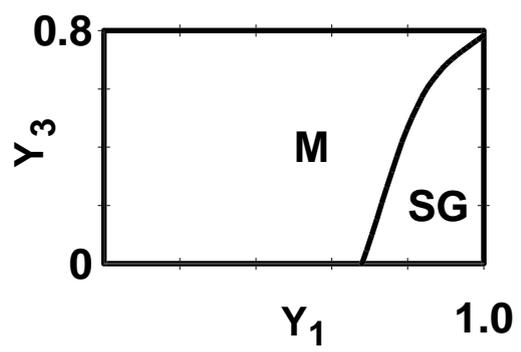
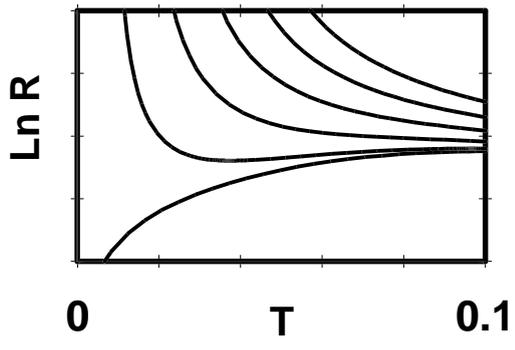
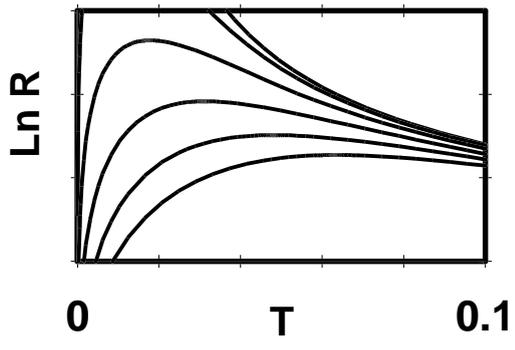
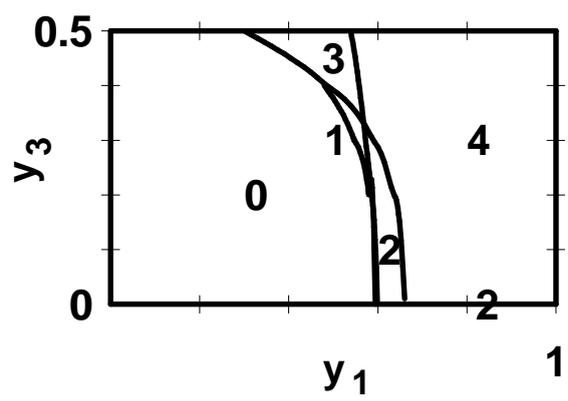
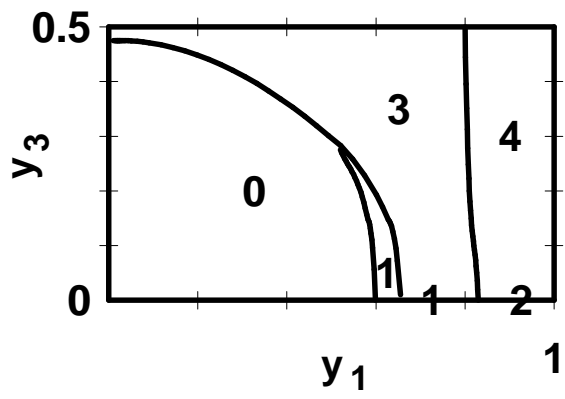
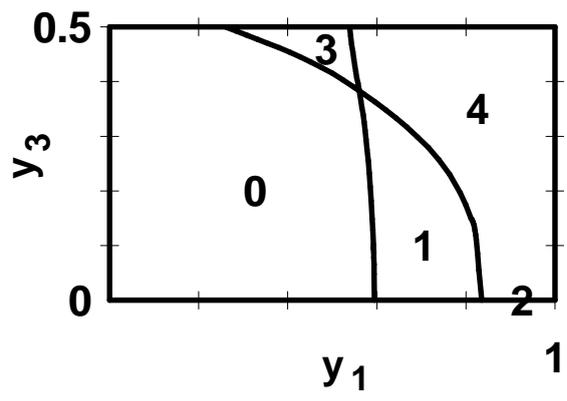
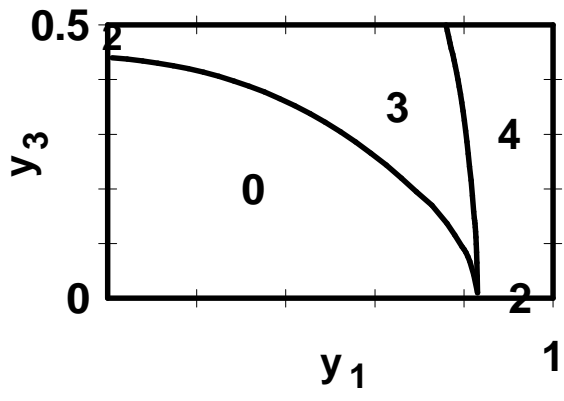
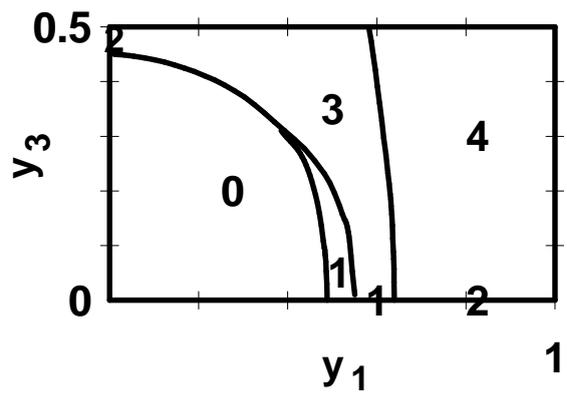
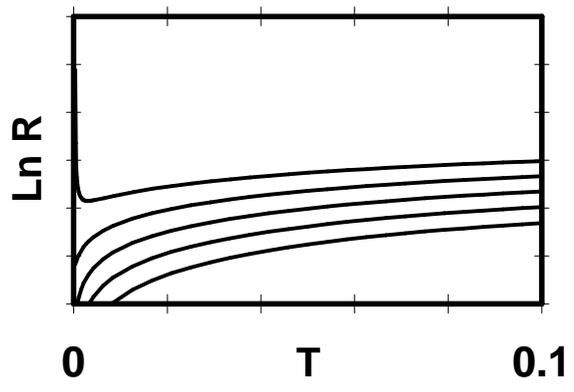
\end{document}